\documentclass[10pt,journal]{IEEEtran}
\usepackage{bm,bbm,amsthm,amsmath,amssymb,amsfonts,mathrsfs}
\usepackage{algorithmic}
\usepackage{algorithm}
\usepackage{array}
\usepackage{textcomp}
\usepackage{stfloats}
\usepackage{url}
\usepackage{verbatim}
\usepackage{cite}
\usepackage{graphicx}
\usepackage{float} 
\usepackage{subfigure}  
\usepackage{multirow,booktabs,threeparttable}
\def\BibTeX{{\rm B\kern-.05em{\sc i\kern-.025em b}\kern-.08em
    T\kern-.1667em\lower.7ex\hbox{E}\kern-.125emX}}
\usepackage{balance}
\newcommand{\eg}{e.g., }
\begin{document}
\title{Stochastic Graph Neural Network-based Value Decomposition for MARL in Internet of Vehicles
}
\author{Baidi Xiao, Rongpeng Li, Fei Wang, Chenghui Peng, Jianjun Wu, Zhifeng Zhao, and Honggang Zhang
    \thanks{B. Xiao, and R. Li are with College of Information Science and Electronic Engineering, Zhejiang University (email: \{xiaobaidi, lirongpeng\}@zju.edu.cn).}
    \thanks{F. Wang, C. Peng, and J. Wu are with Huawei Technologies (email: \{wangfei76, pengchenghui, wujianjun\}@huawei.com).}
    \thanks{Z. Zhao and H. Zhang are with Zhejiang Lab as well as Zhejiang University (email: \{zhaozf, honggangzhang\}@zhejianglab.com).}
    \thanks{Part of the paper has been accepted by IEEE VTC 2023-Spring \cite{SVMIX}.}
}

\maketitle

\begin{abstract}
    Autonomous driving has witnessed incredible advances in the past several decades, while Multi-Agent Reinforcement Learning (MARL) promises to satisfy the essential need of autonomous vehicle control in a wireless connected vehicle networks. In MARL, how to effectively decompose a global feedback into the relative contributions of individual agents belongs to one of the most fundamental problems. However, the environment volatility due to vehicle movement and wireless disturbance could significantly shape time-varying topological relationships among agents, thus making the Value Decomposition (VD) challenging. Therefore, in order to cope with this annoying volatility, it becomes imperative to design a dynamic VD framework. Hence, in this paper, we propose a novel Stochastic VMIX (SVMIX) methodology by taking account of dynamic topological features during the VD and incorporating the corresponding components into a multi-agent actor-critic architecture. In particular, Stochastic Graph Neural Network (SGNN) is leveraged to effectively capture underlying dynamics in topological features and improve the flexibility of VD against the environment volatility. Finally, the superiority of SVMIX is verified through extensive simulations.
\end{abstract}

\section{Introduction}
\label{sec:Intro}
In recent years, the unprecedented development of artificial intelligence (AI) brings the self-driving vehicles \cite{MP, Vision, obstacle} into the spotlight, and these intelligent vehicles form an Internet of Vehicles (IoV).
Equipped with the capability for environmental perception \cite{AutoDriving}, the vehicles try to respond to the observation of the surroundings and find an optimal trajectory between two sites \cite{Ast, RRT} with calibrated self-decision control and traffic congestion avoidance \cite{IDM, V2X}. Additionally, there emerges some research interest towards traffic signal control \cite{Signal} or fleet control \cite{Fleet} in IoV as well. Nevertheless, as for distributed deployment of intelligent vehicles in IoV, such intricate cases (e.g., traffic control) require them to collaborate on top of reliable communication, which further constitutes a Multi-Agent System (MAS) and catalyses the research progress in a myriad of related studies (\eg traffic control and prediction) \cite{NA, TC}. In particular, Multi-Agent Reinforcement Learning (MARL), which incorporates Deep Reinforcement Learning (DRL) into MAS, emerges by astutely learning through the trial-and-error interaction between multiples agents (i.e., vehicles) and the complicated IoV environment, and promises to yield smart policies for the formulated Markov Decision Process (MDP).\par

Traditionally, in Independent Q-Learning (\texttt{IQL}) \cite{IQL}, one typical kind of MARL, each individual agent learns its policy by regarding other agents as parts of the environment, and often experiences a non-stationary environment, as the agents always update their policies independently during the learning. Consequently, the non-stationary issue impedes the optimization of agents' policies. Instead, mutual communication shall be considered for cooperative agents to reach holistic consistency. On the other hand, it is natural for MAS to only observe a global reward from the environment. Thus it becomes essential to tackle with the agent heterogeneity and reward assignment issue. \cite{COMA}.\par
In recent years, algorithms with Centralized Training and Decentralized Execution (CTDE) have become the centerpiece of MARL as they can somewhat handle the aforementioned two problems of \texttt{IQL} (i.e., non-stationary learning, and node heterogeneity). As its name implies, CTDE can be divided into the training phase and execution phase. In particular, in the training phase, agents can implicitly observe the global information of the environment in a centralized manner, so as to guarantee the communication among agents and tackle the non-stationary problem. Subsequently, in the decentralized execution phase, each agent capably makes decisions based on its local observation only. Typical examples of CTDE include \texttt{MADDPG}, \texttt{COMA}, etc \cite{COMA,MADDPG}.
Nevertheless, as the number of agents increases, the centralized state-action value function in CTDE algorithms such as \texttt{COMA} suffers from an exponential increase of the action space as well as the awful growth of computational complexity. Therefore, the Value Decomposition (VD)-based algorithms \cite{QMIX,DOP,VMAC,MGAN} are proposed to decompose the centralized value function into individual value functions, and attempt to learn an implicit representation for individual reward, thus decreasing the computational complexity in CTDE. 
Regretfully, these methods mostly concentrate on learning from agents with the invariant communication topology \cite{Neigh,MGAN}, but flounder under dynamic environments entailing uncertainty and perturbation due to fluctuating connections and dynamically changing topologies.\par
On the other hand, the advent of Graph Neural Network (GNN) \cite{GNN} makes it possible to capture the underlying topological features of vehicle-formed graphs. For example, the spectral-based GNNs \cite{GCN} can filter the external noise and extract features from the transformed frequency domain of signals. However, most of the spectral-based GNNs concentrate on the deterministic graphs during the training phase and ignore the time-varying underlying topologies with internal perturbations in some scenarios (\eg the possibly disconnected communication links due to the vehicle mobility and limited communication ranges in traffic control). Therefore, in order to capture dynamic topological features and better adapt to graph perturbations, it's reasonable to re-design the VD-based method, so as to take account of the graph
stochasticity during the training phase \cite{RES}. \par
In this paper, we focus on dealing with the self-decision control in IoV for intelligent traffic control. On top of the designed mixing network \texttt{VMIX}\cite{QMIX}, we propose the brand-new Stochastic VMIX (\texttt{SVMIX}) MARL algorithm. Compared to previous works in \cite{QMIX}, \cite{VMAC} and \cite{MGAN}, \texttt{SVMIX} adopts Stochastic Graph Neural Network (\texttt{SGNN}) \cite{SGNN} to capture the dynamic topological features of the time-varying graphs and further decompose the value function. Besides, in order to tackle with the continuous action control, \texttt{SVMIX} successfully incorporates the \texttt{VMIX} module into the Proximal Policy Optimization (\texttt{PPO})-based CTDE architecture \cite{PPO}. The contribution of this paper can be summarized as follows.\par
\begin{itemize}
    \item We incorporate \texttt{SGNN} in the VD-based MARL method, which intentionally aggregates the information of agents through different randomly sampled graphs to imitate the practically dynamic connectivity of vehicles and therefore endows vehicles with the capability to resist the environment disturbance. 
    \item We theoretically analyze the role that \texttt{SGNN} plays in the MARL method and manifest the importance and effectiveness of capturing the dynamical topological features to approximate the feasible decomposed solution for the individual value functions of each agent.
    \item We propose a novel MARL method (i.e., \texttt{SVMIX}), which applies \texttt{SGNN}-based VD on top of \texttt{PPO} \cite{PPO} (a kind of RL algorithm within the actor-critic framework), so as to capture the topological feature of the underlying relationship of the agents through a stochastic graph.
    \item We verify the performance of \texttt{SVMIX} on extensive simulation settings. Compared with several benchmark MARL methods (i.e., \texttt{QMIX} \cite{QMIX}, \texttt{MGAN} \cite{MGAN} and \texttt{FIRL} \cite{FIRL}), \texttt{SVMIX} demonstrates its superiority and stability in terms of the convergence rate, system utility and communications overheads.
\end{itemize} \par
The reminder of the paper is organized as follows. Section \ref{sec:related} briefly introduces the related works. Section \ref{sec:background} presents the necessary background and formulated the system model. Section \ref{sec:algorithm} describes the implementation details of the proposed algorithm. In Section \ref{sec:experiment}, we introduce the experimental settings and discuss the related simulation results. Section \ref{sec:conclusion} concludes the paper as well as make a discussion about future works.\par

\section{Related Works}
\label{sec:related}
The traditional physics-based models\cite{CACC, Frequency-domain} for autonomous vehicle control are data-efficient as there are only a few parameters to calibrate with mathematically tractability \cite{Physics-based}. However, these models lie on strong assumptions on traffic scenarios and cannot quickly adapt to the protean environment in practice. Instead, AI has been widely adopted in the field of IoV, and DRL belongs to one of the most fundamental methods \cite{DRL}, taking the observation (e.g., the extracted features from the perception components and the information of other vehicles) as input and making self-decision control (e.g., on the acceleration and direction) the same as the classic controller. Besides, deep learning methods like Convolutional Neural Networks (CNNs) and Recurrent Neural Networks (RNNs) have also made great progress in vision-based detection and prediction-based control \cite{CNN-RNN}. Meanwhile, through learning from the driving data of human beings, imitation learning is able to generate a policy similar to the style of human driving as well\cite{Imitation}. \par
Furthermore, the MARL method stands out for the MAS in IoV where the communication \& collaboration of multiple agents is essential. Notably, a direct application such as \texttt{IQL} faces non-stationary IoV environment while a conventional MARL method encounters reward assignment issues \cite{COMA,IQL}. Instead, CTDE has become one of the most popular paradigms. For example, Refs. \cite{QMIX, DOP, VMAC, MGAN} apply VD and decompose the joint value function into individual value functions corresponding to each agent. These methods somewhat deal with the reward assignment with reduced computational costs. Within the actor-critic framework, Refs. \cite{MADDPG, COMA} utilize a centralized critic for centralized training as well as multiple actors for decentralized execution. In addition, some researchers focus on the communication between agents. Ref. \cite{RIAL} assumes the availability of the communication of all the agents while Ref. \cite{ATOC} relies on the local communication for agents within a certain proximity. \cite{Neigh} further limits the communication to allow each agent to observe the information of its neighbours only. To tackle the exponentially growing action space, Ref. \cite{Mean_Field} applies the Mean Field Theory (MFT) to treat the other agents as an ensemble for every agent, thus significantly reducing the computation cost. \par 
Towards MARL-enabled autonomous vehicle control, Ref. \cite{IPPO} assigns an individual modified \texttt{PPO} agent as the DRL controller for both vehicles and traffic lights, so as to make the traffic lights collaborate with the autonomous vehicles. Specially, the agent in \cite{IPPO} observes the local information independently. Ref. \cite{DIAL} improves the algorithm of \cite{RIAL} through the more effective communication among agents with message approximation and verifies its proposed algorithm in a highway autonomous driving scenario.
Ref. \cite{Games} assumes that each agent can take the other agents' policies into account, and adopts a Markov Game-based DRL. Benefiting from the information exchange among vehicles, Ref. \cite{FIRL} incorporates a consensus algorithm with federated learning on the basis of a CTDE structure. Similarly, Ref. \cite{GCQ} leverages GNN to aggregate the node feature of each vehicle based on the communication graph in IoV and learns a joint policy via a centralized DRL agent. Nevertheless, the aforementioned researches seldom shed light on the dynamic topological connection of the vehicles. 

\section{Preliminaries and System Model}
\label{sec:background}
In this section, we briefly introduce the related knowledge of MDP and VD-based methods, and present the formulated system model to apply MARL for the autonomous vehicle control.
\subsection{Preliminaries}
Generally, an MARL task is formulated as a Decentralized Partially Observable Markov Decision Process (Dec-POMDP) \cite{DecPOMDP}, which is defined as the tuple $\langle\mathcal{I}, \mathcal{S}, \mathcal{A}, \mathcal{P}, \Omega, \mathcal{R}, \mathcal{O}, \gamma\rangle$. $\mathcal{I}$ represents the set of $N$ agents, $\mathcal{S}$ is the state space, $\mathcal{A}$ is the action space for a single agent and $\boldsymbol{\mathcal{A}}:=\mathcal{A}^N$ is the joint action space. The joint action $\boldsymbol{a}=\{{a}^{(1)}, {a}^{(2)}, \cdots, {a}^{(N)}\}$ taken at the current state $s$ results in the next state $s'$ through a transition of environment according to the transition function $\mathcal{P}(s' \vert s, \boldsymbol{a}):\mathcal{S} \times \mathcal{A} \times \mathcal{S} \rightarrow [0, 1]$. Owing to the scant ability of perception against the colossal environment, agent $i$ gets a local observation ${o}^{(i)} \in {\Omega}$ via the observation function $\mathcal{O}({o}^{(i)} \vert s, i):\mathcal{S} \times \mathcal{I} \times {\Omega} \rightarrow [0, 1]$ instead of $s$ at each time-step. All agents share a global reward function $\mathcal{R}(s, \boldsymbol{a}):\mathcal{S} \times \boldsymbol{\mathcal{A}} \rightarrow \mathbb{R}$ and $\gamma$ denotes the discount factor.\par
Furthermore, in a Dec-POMDP, agent $i$ adopts an action according to its policy $\pi^{(i)}(\cdot \vert {o}^{(i)})$, which denotes the conditional probability distribution of taking action ${a}^{(i)}\in\mathcal{A}$ on ${o}^{(i)}$. $\boldsymbol{a}$ is adopted according to the probability $\boldsymbol{\pi}(\boldsymbol{a} \vert \boldsymbol{o})=\prod_{i=1}^N \pi^{(i)}\left({a}^{(i)} \vert o^{(i)}\right)$ where $\boldsymbol{\pi}$ denotes the joint policy and $\boldsymbol{o}=\{{o}^{(1)},\cdots,{o}^{(N)}\}$ implies the joint observation. In order to coordinate all agents for maximizing the discounted accumulated return $J=\sum_{t=0}^{\infty} \gamma^t \mathcal{R}\left(s_t, \boldsymbol{a}_t\right)$, a joint state-action value function is defined as the expected discounted accumulated return starting from $s_t$ and $\boldsymbol{a}_t$ of time $t$, that is 
\begin{equation}
\label{eq:state-action-value-function}
{Q}^{\boldsymbol{\pi}}(s_t, \boldsymbol{a}_t)=\mathbb{E}_{\mathcal{P}}\left[\sum_{k=t}^{\infty} \gamma^{k-t} \mathcal{R}\left(s_k, \boldsymbol{a}_k\right) \vert s_t, \boldsymbol{a}_t, \boldsymbol{\pi}\right],
\end{equation}
where $\mathbb{E}(\cdot)$ denotes an expectation operator.
It can be naturally observed that the function has an exponential computational complexity due to the joint action $\boldsymbol{a}_t$.\par
To deal with this scalability issue while pursuing the optimal joint policy $\boldsymbol{\pi}^{\ast}=\arg\max\nolimits_{\boldsymbol{\pi}}{Q}^{\boldsymbol{\pi}}(s_t, \boldsymbol{a}_t)$, instead of adopting a joint state-action value function as in \eqref{eq:state-action-value-function}, VD-based methods \cite{QMIX,DOP,VMAC,MGAN} adhibit an individual state-action value function ${Q}^{{\pi}^{(i)}}({o}^{(i)}_t, {a}^{(i)}_t)=\mathbb{E}_{\mathcal{P}}\left[\sum_{k=t}^{\infty} \gamma^{k-t} r_k^{(i)} \vert {o}^{(i)}_t, {a}^{(i)}_t, {\pi}^{(i)}\right]$ for agent $i$, where $r_k^{(i)}$ is a learned implicit representation of the individual reward decomposed from $\mathcal{R}(s_k, \boldsymbol{a}_k)$ for agent $i$. Moreover, the individual state-action value function reflects different contributions among agents. \par
Likewise, the expected $J$ from $s_t$ can also be represented by the joint state value function
\begin{equation}
\label{eq:state-value-function}
{V}^{\boldsymbol{\pi}}(s_t)=\mathbb{E}_{\mathcal{P},\boldsymbol{a}_t\sim\boldsymbol{\pi}(\boldsymbol{\cdot} \vert \boldsymbol{o}_t)}\left[\sum_{k=t}^{\infty} \gamma^{k-t} \mathcal{R}\left(s_k, \boldsymbol{a}_k\right) \vert s_t, \boldsymbol{\pi}\right].
\end{equation}
Exactly the joint state value function ${V}^{\boldsymbol{\pi}}(s_t)$ is a comprehensive evaluation of $J$ at $s_t$, so it is commonly used for calculating the advantage value of different actions at $s_t$ as a baseline \cite{Dueling}. Using VD, we can also represent the set of individual state value functions as ${V}^{{\pi}^{(i)}}({o}^{(i)}_t)=\mathbb{E}_{\mathcal{P}, a_t\sim \pi^{(i)}(\cdot \vert o^{(i)}_t)}\left[\sum_{k=t}^{\infty} \gamma^{k-t} r_k^{(i)} \vert {o}^{(i)}_t, {\pi}^{(i)}\right]$, which implies a more reasonable approximation of the joint value function and speeds up the training.
\subsection{System Model}
\begin{figure}
    \centering
    \includegraphics[width=1.05\linewidth]{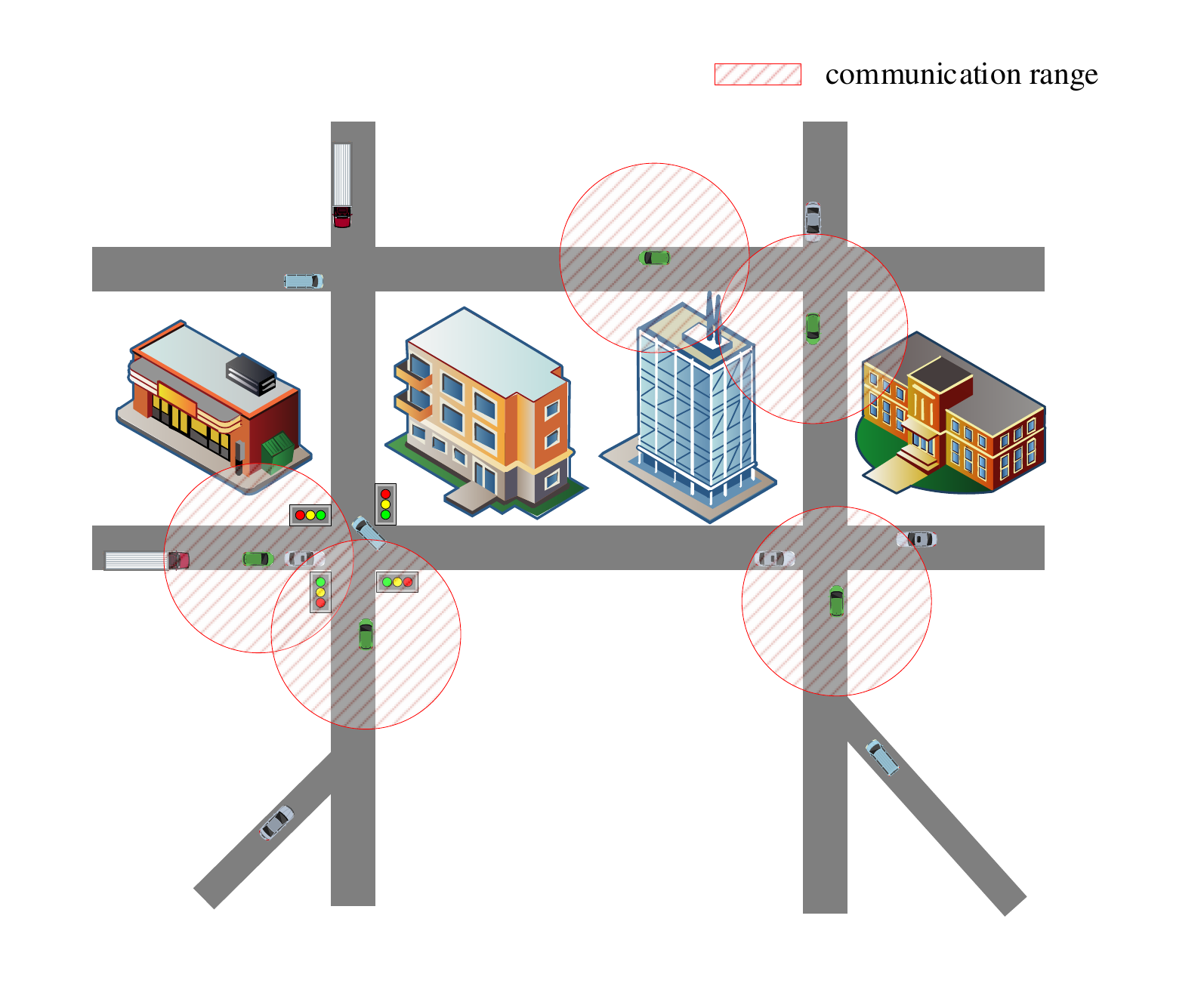}
    \vspace{-1cm}
    \caption{The traffic control scenario with ramps, intersections and vehicles with a constrained communication range.}
    \label{fig1}
\end{figure}

We have proposed a mixed autonomy traffic system model that permits vehicles to flow in and out of road, giving rise to the variable amount of vehicles. At time-step $t$, there are totally $M_t$ vehicles (including $N$ DRL-driven vehicles and $M_t\geq N$) as well as intersections or ramps as depicted in Fig. \ref{fig1}. For vehicle $i$, it has the information such as the velocity $v^{(i)}_t\in \mathbb{R}$ and the position $z^{(i)}_t=(x^{(i)}_t, y^{(i)}_t)\in \mathbb{R}^2$ of itself at time $t$. Meanwhile, $v^{(i)}_t$ is controlled via the acceleration $u^{(i)}_t\in \mathbb{R}$ decided by vehicle $i$ itself. It's also necessary for vehicles to obey some traffic rules. For example, at an intersection, the vehicle has to consider an appropriate passing order as well as control its direction and velocity so as to avoid collisions. 
Based on the above definition, we have the elements of the Dec-POMDP as follows.\par
\paragraph{State and observation} In our scenario, each vehicle can receive the information of other accessible vehicles through Vehicle-to-Vehicle (V2V) connections. Without loss of generality, assuming that vehicle $i$ is only able to communicate with vehicles $j$ and $k$ at time $t$, the observation of vehicle $i$ can be represented as ${o}^{(i)}_t =\{{v^{(i)}_t,z^{(i)}_t,v^{(j)}_t,z^{(j)}_t,v^{(k)}_t,z^{(k)}_t}\}$. The state is represented by $s_t=\{{v^{(1)}_t,z^{(1)}_t,v^{(2)}_t,z^{(2)}_t, \cdots, v^{(M_t)}_t,z^{(M_t)}_t}\}$, which is consisted of the velocities and positions of all the vehicles. Intuitively, the observation ${o}^{(i)}_t$ is a part of $s_t$, which can be mathematically determined by $\mathcal{O}({o}^{(i)}_t \vert s_t, i)$.
\paragraph{Action} At time $t$, vehicle $i$ adopts an action represented as ${a}^{(i)}_t=\{ u^{(i)}_t, q^{(i)}_t\}$ where $q^{(i)}_t$ is an extra action that indicates whether the vehicle changes to another lane or direction. In particular, the $N$ DRL-driven vehicles select their actions according to the learned policies instead of the others with fixed policies.
\paragraph{Reward} The goal of MARL for the vehicle is to maintain a velocity as close to the desired velocity $v_d^{(i)}$ as possible on the basis of no undesirable collisions. Accordingly, the reward $r_t:=\mathcal{R}(s_t, \boldsymbol{a}_t)$ is set as
\begin{equation}
    \label{eq2}
    r_t=
    \begin{cases}\frac{V_d - \sqrt{\sum_{i=1}^{M_t} (v^{(i)}_d-v_t^{(i)})^2}}{V_d+C_1}\\
    \ -\alpha D(z^{(1)}, \cdots, z^{(M_t)}), & \text {if no collision}; \\
             0,                                                                         & \text {otherwise},\end{cases}
\end{equation}

\begin{figure*}
    \centering
    \includegraphics[width=0.8\linewidth]{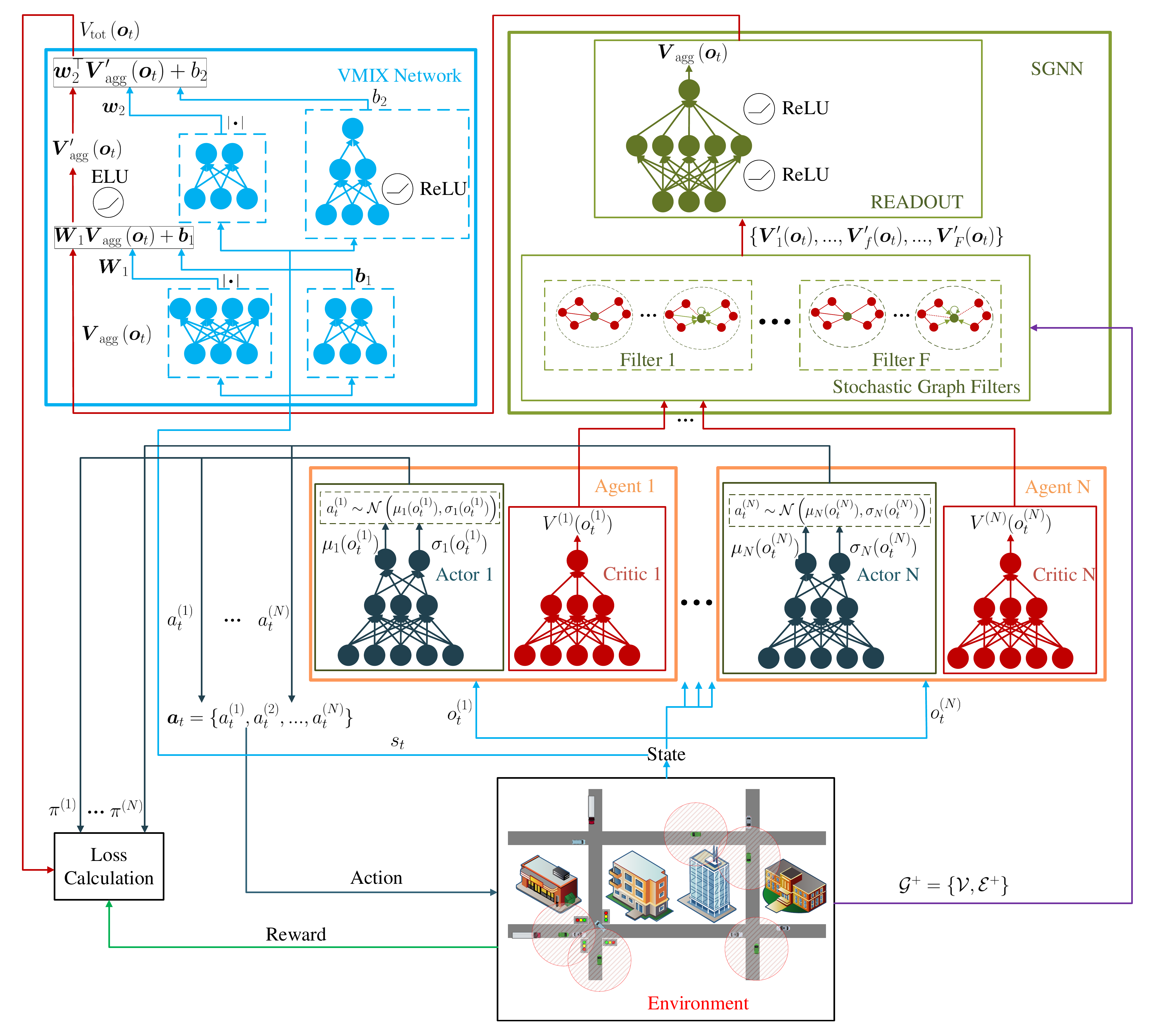}
    \vspace{-0.2cm}
    \caption{The illustration of the \texttt{SVMIX} algorithm for autonomous vehicle control.}
    \label{fig2}
\end{figure*}

\noindent where $V_d=\sqrt{\sum_{i=1}^{M_t} \left( v^{(i)}_d \right)^2}$ and $C_1$ is a very small constant. $D(z^{(1)}, \cdots, z^{(M_t)})$ is the penalty term with a safety coefficient $\alpha$ that hopes the vehicle to hold a safe distance with its preceding vehicle. \eqref{eq2} implies that MARL encourages the vehicle to choose a satisfactory velocity which deviates trivially from $v^{(i)}_d$ while spares no efforts to avoid collisions at time $t$. Notably, all the $M_t$ vehicles participate in the calculation of the reward.\par
On the other hand, the wireless-connected intelligent vehicles in IoV constitute a time-varying undirected graph $\mathcal{G}_t=\{\mathcal{V},\mathcal{E}_t\}$, where each agent is denoted as a vertex and a communication link are regarded as an edge. In other words, $\mathcal{V}=\{1,2,\cdots,N\}$ and an undirected edge in $\mathcal{E}_t$ corresponds to a communication link between two vehicles. It is worth recalling that the mobility of vehicles will frequently change the connections among vehicles due to the communication range and thus induce topological changes. For example, if a vehicle meets another vehicle at an intersection, it's obvious that a strong connection between them could be set up in order to avoid collisions. Additionally, the time-varying amount of vehicles also results in the fluctuation of the number of vertices in $\mathcal{G}_t$. For simplicity and without losing the rationality, we will treat $\mathcal{G}_t$ as a graph with fixed vertices. 
\subsection{Problem Formulation}
Following the system model, we expect the DRL-driven vehicle to maintain a regular speed that close to its desired velocity on the premise of avoiding collisions. Therefore we have proposed a system utility function $U$ as the objective of policy optimization, which can be mathematically formulated as
\begin{equation}
\label{eq:problem_objectives}
    \max_{\boldsymbol{\pi}} \;U=
    \mathbb{E}_{t}\left[r_t \vert \boldsymbol{\pi}\right].   
\end{equation}
\noindent In other words, the agents aim to learn a joint policy $\boldsymbol{\pi}$ under the guidance of maximizing the system utility (i.e., the expected reward). Notably, for CTDE, the optimization of $\boldsymbol{\pi}$ is equivalent to learn the optimal individual policies of agents \cite{QMIX}. Therefore, \eqref{eq:problem_objectives} implies to devise an appropriate VD method for CTDE in volatile environment, so as to derive an individual policy of each agent. 

\section{\textsc{Architecture and Designation of \texttt{\large SVMIX}}}
\label{sec:algorithm}

In this section, we describe the architecture and components of the proposed \texttt{SVMIX} algorithm as depicted in Fig. \ref{fig2}. First, we use \texttt{PPO} as the basis of each individual DRL agent.
Afterwards, in order to decompose the proper individual value functions of the $N$ agents from the total (or joint) value function through the \texttt{VMIX} network, we highlight how to leverage \texttt{SGNN} \cite{SGNN} to effectively capture the topological features from the dynamic graph $\mathcal{G}_t$. \par 
Next we will introduce these essential components (i.e., \texttt{PPO}, \texttt{SGNN} and \texttt{VMIX}) in \texttt{SVMIX} as well as the training procedure. As \texttt{SGNN} is the centerpiece of our algorithm, we also mathematically explain that \texttt{SGNN} can help \texttt{PPO} agents to learn the average optimal solutions.

\subsection{\texttt{PPO}-based RL Agents}
Belonging to one of the most famous Policy Gradient (PG) algorithms extensively used for continuous action control, \texttt{PPO} is composed of an actor network and a critic network, where the former is responsible for taking actions in accordance with its learned policy while the latter produces an approximated (individual) value function. Notably, in this paper, we consider a DRL-driven vehicle and a \texttt{PPO}-based agent is equivalent. \par
As illustrated in the lower part of Fig. \ref{fig2}, for agent $i$, the actor outputs the mean $\mu_i(o^{(i)}_t)$ and the standard deviation $\sigma_i(o^{(i)}_t)$ of a normal distribution $\mathcal{N}(\mu_i(o^{(i)}_t), \sigma_i^2(o^{(i)}_t))$ (i.e. the policy $\pi^{(i)}(\cdot \vert o^{(i)}_t)$) through a Multi-Layer Perception (MLP) taking $o^{(i)}_t$ as input with nonlinear activation functions (omitted from Fig. \ref{fig2} for simplicity). Sequentially, following $\pi^{(i)}(\cdot \vert o^{(i)}_t)$, the \texttt{PPO} agent samples an action $a^{(i)}_t$ from $\mathcal{N}(\mu_i(o^{(i)}_t), \sigma_i^2(o^{(i)}_t))$ at time $t$ for exploration. Besides, the critic approximates the state value function and produces a state value $V^{(i)}(o^{(i)}_t)$ through another MLP. \par
After that, the joint action $\boldsymbol{a}_t$ is taken and the environment correspondingly enters the next state $s_{t+1}$, thus shifting the next joint observation to $\boldsymbol{o}_{t+1}$ and yielding a global reward $r_t$. Notably, during this procedure, to better represent the total value function related to the dynamic graph, individual state values are further processed through \texttt{SGNN}-based feature capturing and aggregation.

\subsection{Topological Feature Capturing and Filtering by \texttt{SGNN}}
As mentioned before, the vehicles can form an undirected graph $\mathcal{G}_t=\{\mathcal{V},\mathcal{E}_t\}$. Thus motivated by the extraordinary achievements of GNN to tackle the topology issues, we propose to apply graph signal processing techniques for feature capturing, by treating each individual state value $V^{(i)}(o^{(i)}_t)$ as the signal of one corresponding vertex in $\mathcal{G}_t$. Furthermore, we argue that \texttt{SGNN} makes a good complement to deal with the dynamic graph in traffic control. \par
The upper-right part of Fig. \ref{fig2} demonstrates that \texttt{SGNN} consists of stochastic graph filters and the $\texttt{READOUT}$ mechanism for output integration. \texttt{SGNN} mainly adopts stochastic graph filters to fit the environment volatility based on the Random Edge Sampling (RES) model \cite{RES} with the underlying graph $\mathcal{G}^+=\{\mathcal{V},\mathcal{E}^+\}$, wherein $\mathcal{E}^+$ encompasses all sets of edges $\mathcal{E}_t$ for all $t$ (i.e., $\mathcal{E}_t \subseteq \mathcal{E}^+, \forall t$). Besides, we define the adjacent matrix ${\bf{A}}\in \mathbb{R}^N\times\mathbb{R}^N$ where an entry ${{\bf A}}_{m,n}$ is nonzero if and only if $(m,n)\in \mathcal{E}^+$, as well as the Laplacian matrix ${\bf L} = \operatorname{diag}({\bf A}{\bf 1}) - {\bf A}$ for $\mathcal{G}^+$. 
In particular, a random sub-graph $\mathcal{G}^+_k=\{{\mathcal{V},\mathcal{E}^+_k}\}$ will be constructed by holding the original vertices of $\mathcal{G}^+$ but sampling the edges from $\mathcal{E}^+$ following a Bernoulli distribution with a success probability $p$, that is,
\begin{equation}
    \label{eq3}
    \Pr\left[(m,n)\in \mathcal{E}^+_k\right]=p, \quad \text {for all } (m,n)\in \mathcal{E}^+.
\end{equation}

\begin{figure}
    \centering
    \includegraphics[width=1\linewidth]{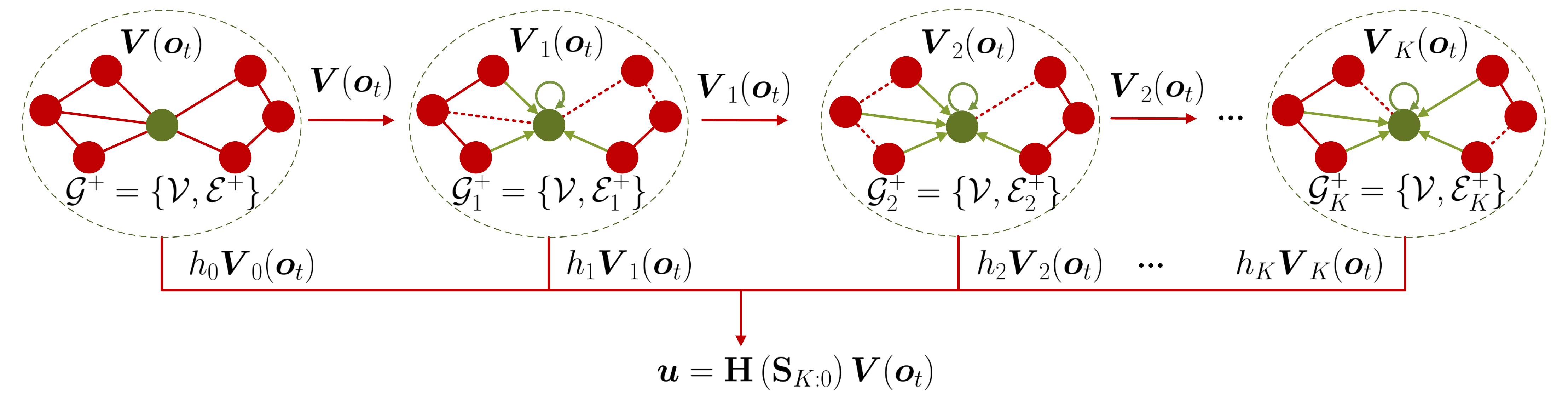}
    \vspace{-0.6cm}
    \caption{A stochastic graph filter of $K$ orders. Here we only show how the information of the green vertex is updated via green edges.}
    \label{fig3}
\vspace{-0.2cm}
\end{figure}

\noindent Through RES, we can emulate dynamic graphs where the communication links (edges) vary frequently. Therefore, to exert the advantage of centralized training, we actually let $\mathcal{G}^+$ be a generalized fixed graph encompassing all potential connections between vertices, so as to ensure enough exploration.\par
In detail, Fig. \ref{fig3} demonstrates the corresponding structure of a $K$-order stochastic graph filter. Based on the input $\boldsymbol{V} (\boldsymbol{o}_t)=\{V^{(1)}(o^{(1)}_t),\cdots,V^{(N)}(o^{(N)}_t)\}\in \mathbb{R}^N$ from \texttt{PPO} agents as well as the stochastic sub-graphs $\mathcal{G}^+_k$ with the adjacent matrix ${\bf A}_k$, the information among vertices recursively diffuses as $\boldsymbol{V}_k(\boldsymbol{o}_t)={\bf S}_k\boldsymbol{V}_{k-1}(\boldsymbol{o}_t), k \in \{1, \cdots, K\}$ with $\boldsymbol{V}_0(\boldsymbol{o}_t) = \boldsymbol{V}(\boldsymbol{o}_t)$ and either ${\bf S}_k={\bf A}_k+{\bf I}$ (where ${\bf I}$ denotes the identity matrix) or ${\bf S}_k={\bf L}_k$. Therefore, having ${\bf S}_0={\bf I}$ and unrolling the recursion, the intermediate signal can be represented as
\begin{equation}
    \label{eq4}
    \begin{aligned}
        \boldsymbol{V}_k(\boldsymbol{o}_t)={\bf S}_k\boldsymbol{V}_{k-1}(\boldsymbol{o}_t)&=({\bf S}_k{\bf S}_{k-1}\cdots{\bf S}_0)\boldsymbol{V}(\boldsymbol{o}_t)\\
        &:={\bf S}_{k:0}\boldsymbol{V}(\boldsymbol{o}_t),
    \end{aligned}
\end{equation}
where ${\bf S}_{k:0}$ is defined as ${\bf S}_{k:0}:={\bf S}_k{\bf S}_{k-1}\cdots{\bf S}_0$. Therefore, $\boldsymbol{V}_k(\boldsymbol{o}_t)$ aggregates the information of $\boldsymbol{V}(\boldsymbol{o}_t)$ through the random sequence of sub-graphs $\mathcal{G}^+_1,\cdots,\mathcal{G}^+_k$. Ultimately, the output $\boldsymbol{u}$ of the $K$-order stochastic graph filter with filter coefficients $h_k$ can be formulated as
\begin{equation}
    \label{eq5}
    \boldsymbol{u} =\sum_{k=0}^{K} h_k\boldsymbol{V}_k(\boldsymbol{o}_t)=\sum_{k=0}^{K} h_k{\bf S}_{k:0}\boldsymbol{V}(\boldsymbol{o}_t):={\bf H}\left({\bf S}_{K:0}\right)\boldsymbol{V}(\boldsymbol{o}_t) .
\end{equation}
Specifically, ${\bf H}\left({\bf S}_{K:0}\right)$ denotes the stochastic graph filter with learnable parameters $h_0,\cdots, h_K$.\par
Furthermore, we can use $F$ parallel stochastic graph filters ${\bf H}_f\left({\bf S}_{K:0}\right)$ with coefficients $h_{fk} \in \mathcal{H}, \forall f \in \{1,\cdots,F\}, k \in \{1,\cdots, K\}$ to process $\boldsymbol{V} (\boldsymbol{o}_t)$, where $\mathcal{H}$ is the set of all the filter coefficients. In particular, for the $f$-th stochastic graph filter, the output can be given on top of \eqref{eq5} as
\begin{equation}
    \label{eq6}
    \begin{aligned}\boldsymbol{V}'_f(\boldsymbol{o}_t)=\sigma\left[\boldsymbol{u}_f\right]&=\sigma\left[{\bf H}_f\left({\bf S}_{K:0}\right)\boldsymbol{V} (\boldsymbol{o}_t)\right]\\
        &=\sigma\left[\sum_{k=0}^{K}h_{fk} {\bf S}_{f,k:0}\boldsymbol{V} (\boldsymbol{o}_t)\right],
        \hspace{-1mm}
    \end{aligned}
\end{equation}
where $\sigma(\cdot)$ is the nonlinear activation function ReLU. \par
Subsequently, the mechanism $\texttt{READOUT}:\mathbb{R}^N\times\mathbb{R}^F \rightarrow \mathbb{R}^N$ is implemented to integrate the outputs of the stochastic graph filters through a two-layer MLP as
\begin{equation}
    \label{eq7}
    \boldsymbol{V}_{\rm agg}(\boldsymbol{o}_t)=\texttt{\large READOUT}(\{\boldsymbol{V}'_1(\boldsymbol{o}_t),\cdots,\boldsymbol{V}'_F(\boldsymbol{o}_t)\}).
\end{equation}\par
Finally, \texttt{SGNN} captures the dynamic topological features through stochastic graph filters and random sub-graphs generated by RES with learnable filter coefficients $\mathcal{H}$. That is, the filtered state values $\boldsymbol{V}_{\rm agg}(\boldsymbol{o}_t)$ involve the dynamic topological features. Notably, consistent with methodology of CTDE, \texttt{SGNN} requires the information in every vertex as input in the centralized training phase, while it can be neglected at the decentralized execution phase.\par

\subsection{The Role of \texttt{SGNN}}
Intuitively, RES in \texttt{SGNN} brings substantial uncertainty to capture underlying topological features. Therefore, it's necessary to clarify the role of \texttt{SGNN} in \texttt{SVMIX} with theoretical analysis. 
Concentrating on the stochastic graph filter defined in \eqref{eq5} on the basis of RES, for a given graph $\mathcal{G}^+$ and an input $\boldsymbol{V}(\boldsymbol{o}_t)$, the expected output can be given by
\begin{equation}
    \label{expected_output}
    \mathbb{E}\left[\boldsymbol{u}\right] =\mathbb{E}\left[{\bf H}\left({\bf S}_{K:0}\right)\boldsymbol{V}(\boldsymbol{o}_t)\right]=\mathbb{E}\left[\sum_{k=0}^{K} h_k{\bf S}_{k:0}\right]\boldsymbol{V}(\boldsymbol{o}_t).
\end{equation}
Let ${\bf S}_k=\overline{\bf{S}}+\Delta_k$ where $\overline{\bf{S}}=\mathbb{E}\left[{\bf S}_k\right] $ and $\Delta$ represents the error matrix, we further get $\mathbb{E}\left[\Delta_k\right] = 0$ and 
\begin{equation}
    \label{expected_shift}
    \begin{aligned}
    \mathbb{E}\left[\sum_{k=0}^{K} h_k{\bf S}_{k:0}\right]&=\sum_{k=0}^{K} h_k\mathbb{E}\left[{\bf S}_k{\bf S}_{k-1}\cdots{\bf S}_0\right] \\
    &=\sum_{k=1}^{K} h_k\mathbb{E}\left[\left(\overline{\bf{S}}+\Delta_k\right)\cdots\left(\overline{\bf{S}}+\Delta_1\right)\right] + h_0\textbf{I} \\
    &=\sum_{k=1}^{K} h_k\overline{\bf{S}}^k + h_0\textbf{I}
    \end{aligned}
\end{equation}
as $\mathbb{E}\left[\Delta_m\Delta_n\right]=\mathbb{E}\left[\Delta_m\right]\mathbb{E}\left[\Delta_n\right]=0$ if $m\neq n$ given the mutual independence between the samplings.\par
\begin{itemize}
    \item[1)] 
    If ${\bf S}_k={\bf L}_k$, for an entry $\overline{\bf S}_{m, n}$ in $\overline{\bf S}$, we have
\begin{equation}
    \label{expected_L}
    \overline{\bf S}_{m, n}=
    \begin{cases} -p{\bf A}_{m, n}, & m \neq n; \\
             pd_m, & \text {else},\end{cases}
\end{equation}
where $d_m$ is represented by the degree of vertex $m$. So $\overline{\bf S}=p{\bf L}$, and finally we get $\mathbb{E}\left[\boldsymbol{u}\right]=\left(\sum_{k=1}^{K} h_k p^k {\bf L}^k + h_0\textbf{I}\right)\boldsymbol{V}(\boldsymbol{o}_t)$.\par
As ${\bf L}$ is a real symmetry matrix, it can be transformed into ${\bf L}={\bf U}\Sigma{\bf U}^T$ through eigendecomposition where $\Sigma=\operatorname{diag}\left(\lambda_1, \lambda_2,\cdot,\lambda_N\right)$ with $\lambda_1\geq \lambda_2\geq\cdots\geq\lambda_N$. Notably, $\lambda_N \equiv 0$ as ${\bf L}$ always has an eigenvector ${\bf 1}$ corresponding to the eigenvalue $0$, and the algebraic multiplicity of $\lambda_N$ equals to the number of connected components in $\mathcal{G}^+$ \cite{eigen}. Considering the influence of filtering coefficients $h_k$ and probability $p$, $\sum_{k=1}^{K} h_k p^k {\bf L}^k + h_0\textbf{I}$ will be a full-rank matrix (e.g., if $h_k>0, \forall k=0,1,\cdots,K$, the eigenvalues will be $\tilde{\lambda}_1\geq\tilde{\lambda}_2\geq\cdots\geq\tilde{\lambda}_N=h_0>0$).
    \item[2)] 
    If ${\bf S}_k={\bf A}_k+{\bf I}$, we have $\overline{\bf S}_{m, n}$ as
    \begin{equation}
    \label{expected_A}
    \overline{\bf{S}}_{m, n} = \mathbb{E}\left[{\bf A}_k+{\bf I}\right]_{m, n}=
    \begin{cases} p{\bf A}_{m, n}, & m \neq n; \\
             1, & \text {else}.\end{cases}
    \end{equation}
    We can also express the real symmetry matrix as $\overline{\bf S}={\bf U}\Sigma{\bf U}^T$ via eigendecomposition where $\Sigma=\operatorname{diag}\left(\lambda_1, \lambda_2,\cdot,\lambda_N\right)$. Assuming that $\mathcal{G}^+$ is a fully connected graph, there will be $\lambda_1=1+(N-1)p$ and $\lambda_2=\lambda_3=\cdots=\lambda_N=1-p$. Thus, choosing a feasible $p$ and taking $h_k$ into account, $\sum_{k=1}^{K} h_k\overline{\bf{S}}^k + h_0\textbf{I}$ will be a full-rank matrix as well.
\end{itemize} \par
Based on the above analysis, we have proved that $\mathbb{E}\left[{\bf H}\left({\bf S}_{K:0}\right)\right]$ can be full-rank with appropriate filtering coefficients $h_k$ and $p$. Hence, for a given expected output $\mathbb{E}\left[\boldsymbol{u}\right]$, there must be a nontrivial solution $\boldsymbol{V}(\boldsymbol{o}_t)$ that satisfies \eqref{expected_output} with learned $\mathcal{H}$.\par
    Together with \eqref{eq6} and \eqref{eq7} which show the relationship between $\boldsymbol{u}$ and $\boldsymbol{V}_{\rm agg}(\boldsymbol{o}_t)$, it becomes reasonable to use $\texttt{SGNN}$ for feature capturing because $\boldsymbol{V}(\boldsymbol{o}_t)$ can finally be mapped to ${V}_{\text{agg}}^{(i)}(\boldsymbol{o}_t)$ where 
    \begin{equation}
    \label{true_valuefunction}
    \begin{aligned}
    {V}_{\text{agg}}^{(i)}(\boldsymbol{o}_t) &=\left[\boldsymbol{V}_{\rm agg}(\boldsymbol{o}_t)\right]_i \\
    &=\mathbb{E}_{\mathcal{P}, \boldsymbol{a}_t\sim \boldsymbol\pi(\cdot \vert \boldsymbol{o}_t)}\left[\sum_{k=t}^{\infty} \gamma^{k-t} r_k^{(i)} \vert \boldsymbol{o}_t, \boldsymbol{\pi}\right].
    \end{aligned}
    \end{equation}
    In other words, the individual value functions take the observations and policies of other agents as well as the dynamics of topology into account in an effort to handle the nonstationarity through stochastic graph filters after sufficient training. \par
 By the way, as $\mathcal{G}^+$ is generalized for all possible $\mathcal{G}_t$ through RES during training, \texttt{SGNN} is able to explore the topologies that may appear, thus learning stable graph signal filters. Therefore, \texttt{SVMIX} ultimately obtain the capability of anti-disturbance.\par

\subsection{Value Decomposition with \texttt{VMIX}}
Similar to the mixing network in \texttt{QMIX} \cite{QMIX}, the \texttt{VMIX} network leverages the filtered signals from \texttt{SGNN} and evaluates the contribution of each agent to generate the aggregated total state value $V_{\rm tot}\left(\boldsymbol{o}_t\right)$ on the basis of the global state $s_t$.
\texttt{VMIX} consists of multiple MLPs which take $s_t$ as input and output the weights and biases for linear transformations. Finally, the aggregated state value can be formulated as
\begin{equation}
    \label{eq9}
    V_{\rm tot}(\boldsymbol{o}_t)=\boldsymbol{w}_2^{\top}\boldsymbol{V}'_{\rm agg}\left(\boldsymbol{o}_t\right)+b_2,
\end{equation}
where $\boldsymbol{V}'_{\rm agg}(\boldsymbol{o}_t)=f\left[\boldsymbol{W}_1\boldsymbol{V}_{\rm agg}\left(\boldsymbol{o}_t\right)+\boldsymbol{b}_1\right]$. Besides, $\boldsymbol{W}_1 \in \mathbb{R}^C \times \mathbb{R}^N$, $\boldsymbol{b}_1 \in \mathbb{R}^C$, $\boldsymbol{w}_2 \in \mathbb{R}^C$, $b_2 \in \mathbb{R}$ are the generated hyperparameters indicating non-negative weights and biases from MLPs taking $s_t$ as input while the superscript $\top$ means the transpose. Besides, $f(\cdot)$ denotes the nonlinear ELU function and enables \texttt{VMIX} to produce a nonlinear total value function $V_{\rm tot}\left(\boldsymbol{\cdot}\right)$. \par
Thus via \texttt{VMIX}, the individual state values $\boldsymbol{V}_{\rm agg}(\boldsymbol{o}_t)$ with captured topological features from \texttt{SGNN} are further integrated into the total state value $V_{\rm tot}(\boldsymbol{o}_t)$ by nonlinear transformation, where the weights and biases learn the contribution of each agent under the guidance of the global state $s_t$. Conversely, $V^{(i)}(o_t^{(i)})$ is decomposed from $V_{\rm tot}(\boldsymbol{o}_t)$ via $\boldsymbol{V}_{\rm agg}(\boldsymbol{o}_t)$. Furthermore, through applying gradient descent and back propagation, the parameters of each critic will be updated and thus an appropriate $V^{(i)}(\cdot)$ will be learned.

\subsection{The Training of \texttt{SVMIX}}

The \texttt{SVMIX} network aims to learn a joint policy $\boldsymbol{\pi}(\boldsymbol{\cdot} \vert \boldsymbol{o}_t;{\theta})=\prod_{i=1}^N \pi^{(i)}\left(\cdot \vert o^{(i)}_t;{\theta}^{(i)}\right)$ (i.e., the normal distribution $\mathcal{N}(\mu_i(o^{(i)}_t), \sigma_i^2(o^{(i)}_t))$ for $i=1,\cdots,N$) and a total state value function $V_{\rm tot}(\boldsymbol{\cdot};{\phi,\eta,\omega})$, where ${\theta}=\langle\theta^{(1)},\cdots,\theta^{(N)}\rangle$ and ${\phi}=\langle\phi^{(1)},\cdots,\phi^{(N)}\rangle$ are the learnable parameters of actors and critics respectively. Besides, $\eta$ and $\omega$ indicate the parameters of \texttt{SGNN} and \texttt{VMIX}, respectively. According to $\boldsymbol{\pi}(\boldsymbol{\cdot} \vert \boldsymbol{o}_t;{\theta})$, the joint action $\boldsymbol{a}_t$ is adopted. In particular, $\boldsymbol{a}_t$ indicates the taken actions of vehicles at time $t$ while $\boldsymbol{\pi}(\boldsymbol{a}_t \vert \boldsymbol{o}_t;{\theta})$ and $V_{\rm tot}(\boldsymbol{o}_t;{\phi,\eta,\omega})$ participates in the calculation of loss functions with the global reward $r_t$. Inspired by \cite{PPO}, the loss functions are defined as
\begin{align}
     & L(\theta) =-{\mathbb{E}}_t\left[\min \left(\rho_{t;\theta} {A}_{t;\phi,\eta,\omega}, \operatorname{clip}\left(\rho_{t;\theta},\epsilon\right) A_{t;\phi,\eta,\omega}\right)\right],\label{eq10}\\
     & L(\phi,\eta,\omega) =\frac{1}{2} A_{t;\phi,\eta,\omega}^2\label{eq11},
\end{align}
where \begin{equation}
    \label{eq12}
    \quad\rho_{t;\theta} = \frac{\boldsymbol{\pi}(\boldsymbol{a}_t \vert \boldsymbol{o}_t;{\theta})}{\boldsymbol{\pi}(\boldsymbol{a}_t \vert \boldsymbol{o}_t;{\theta}_{\rm old})}
\end{equation}
and
\begin{equation}
    \label{eq13}
    \quad{A}_{t;\phi,\eta,\omega} = \sum_{k=t}^T \gamma^{k-t} r_k - V_{\rm tot}(\boldsymbol{o}_t;{\phi,\eta,\omega}).
\end{equation}
Here $T$ is the length of the episode, the clip function $\operatorname{clip}(\cdot)$ removes the incentive for moving the ratio $\rho_{t}$ outside of the interval $[1-\epsilon, 1+\epsilon]$ and ${A}_t$ is the advantage value that evaluates the current policy in the form of Monte-Carlo error. Notably, for training with batches that include time-related samples, \eqref{eq13} will further be slightly modified as
\begin{equation}
    \begin{aligned}
        \label{advantage_in_batch}
        {A}_{j;\phi,\eta,\omega} = \sum_{k=j}^{\vert \mathcal{B} \vert} \gamma^{k-j} r_k &+ \gamma^{ \vert \mathcal{B} \vert -j}V_{\rm tot}(\boldsymbol{o}_{\vert \mathcal{B} \vert + 1};\phi,\eta,\omega)\\
        &- V_{\rm tot}(\boldsymbol{o}_j;{\phi,\eta,\omega}),
    \end{aligned}
\end{equation}

\begin{algorithm}[t]
    \label{Algorithm1}
	\caption{The Training of \texttt{SVMIX} Algorithm}
	\begin{algorithmic}[1]
		\STATE Initialize the underlying graph $\mathcal{G}^+=\{\mathcal{V},\mathcal{E}^+\}$, the number of agents $N$, the batch size $N_{\rm batch}$, the length of episodes $T$, the number of \texttt{PPO} epochs $N_{\rm epoch}$, discount factor $\gamma$ and constant $\epsilon$;
        \STATE Initialize the actor and critic networks in \texttt{PPO} agents, the \texttt{SGNN} network and the \texttt{VMIX} network with random parameters $\theta$, $\phi$, $\eta$ and $\omega$ respectively;
        \STATE Initialize the batch $\mathcal{B}\leftarrow\varnothing$ and the batch sample counter as $\texttt{counter} \leftarrow 0$;
		\FOR{every episode}
            \STATE Initialize the ending flag of the episode as $\texttt{done} \leftarrow 0$;
            \FOR{$t\leftarrow$  1 to $T$ do}
                \STATE Obtain the global state $s_t$ and the joint observation $\boldsymbol{o}_t=\{{o}^{(1)}_t,\cdots,{o}^{(N)}_t\}$ from the environment;
                \STATE Each agent generates the mean $\mu_i(o^{(i)}_t;\theta^{(i)})$ and the standard deviation $\sigma_i(o_t^{(i)};\theta^{(i)})$ of the normal distribution $\mathcal{N}(\mu_i(o^{(i)}_t;\theta^{(i)}), \sigma_i^2(o^{(i)}_t;\theta^{(i)}))$;
                \STATE Each agent chooses the sampled action $a^{(i)}_t\sim\mathcal{N}(\mu_i(o^{(i)}_t;\theta^{(i)}), \sigma_i^2(o^{(i)}_t;\theta^{(i)}))$ ;
                \STATE Obtain the reward $r_t$, $s_{t+1}$  and $\boldsymbol{o}_{t+1}$ at the end of $t$;
                \STATE Store the tuple $\langle s_t,\boldsymbol{o}_t,\boldsymbol{a}_t,r_t,s_{t+1},\boldsymbol{o}_{t+1}\rangle$ in $\mathcal{B}$ in order of time and set $\texttt{counter} \leftarrow \texttt{counter} + 1$;
                
                \IF{$t=T$} 
                \STATE $\texttt{done} \leftarrow 1$;
                \ENDIF                
                
                \IF{$\texttt{counter}=N_{\rm batch}$ or $\texttt{done} = 1$}
                    \STATE Clone $\theta_{\rm old} \leftarrow \theta$;
                    \FOR{$n_{\rm{epoch}}\leftarrow$  1 to $N_{\rm epoch}$ do}
                        \FOR{samples in $\mathcal{B}$}
                            \STATE Obtain $\boldsymbol{\pi}(\boldsymbol{a}_j \vert \boldsymbol{o}_j;{\theta})$ and $\boldsymbol{\pi}(\boldsymbol{a}_j \vert \boldsymbol{o}_j;{\theta}_{\rm old})$ corresponding to the distributions from the actors and calculate $\rho_{j;\theta}$ by (\ref{eq12});
                            \STATE Obtain the approximated total state value $V_{\rm tot}(\boldsymbol{o}_j;\phi,\eta,\omega)$ through the critic networks, \texttt{SGNN} and \texttt{VMIX} network sequentially in (\ref{eq6})-(\ref{eq9});
                            \STATE Calculate ${A}_{j;\phi,\eta,\omega} \leftarrow \sum_{k=j}^{\vert \mathcal{B} \vert} \gamma^{k-j} r_k + (1-\texttt{done})*\gamma^{ \vert \mathcal{B} \vert -j}V_{\rm tot}(\boldsymbol{o}_{\vert \mathcal{B} \vert + 1};\phi,\eta,\omega)- V_{\rm tot}(\boldsymbol{o}_j;{\phi,\eta,\omega})$ ;
                        \ENDFOR
                        \STATE Update $\theta$ and $\phi,\eta,\omega$ according to (\ref{eq10}) and (\ref{eq11}) separately via batch gradient descent;
                    \ENDFOR
                     \STATE Initialize $\mathcal{B}\leftarrow\varnothing$ and $\texttt{counter} \leftarrow 0$;
                \ENDIF
            \ENDFOR
		\ENDFOR
	\end{algorithmic}
\end{algorithm}

where $\vert \mathcal{B} \vert$ is the batch size and $j=\{1,\cdots,\vert \mathcal{B} \vert\}$ is the index of samples. If the episode ends, \eqref{advantage_in_batch} will degenerate into the form of (\ref{eq13}) as the approximation of subsequent rewards is no longer needed.
Finally, ${\theta}$ and $\phi,\eta,\omega$ will be updated through gradient descent to minimize \eqref{eq10} and \eqref{eq11} separately. \par
Substantially, $\texttt{SVMIX}$ decomposes ${V}^{(i)}({o}^{(i)}_t)$ from $V_{\rm tot}(\boldsymbol{o}_t)$ through gradient descent on the minimization of \eqref{eq11} via \texttt{VMIX}. In particular, with the aid of \texttt{SGNN}, a feasible mapping from $\boldsymbol{V}(\boldsymbol{o}_t)$ to ${V}_{\text{agg}}^{(i)}(\boldsymbol{o}_t)$ which takes the dynamics of topology and the influence from other agents into account is learned first. Afterwards, a satisfied joint policy will be learned according to \eqref{eq10}. Additionally, the RES model in \texttt{SGNN} enhances the anti-disturbance ability of $\texttt{SVMIX}$ and the exploration ability by capturing the dynamic topological feature of $\mathcal{G}_t$ through $\mathcal{H}$ and $p$. In other words, for all possible inputs $\boldsymbol{V}(\boldsymbol{o}_t)$ related to a specific topological connection, $\texttt{SGNN}$ helps to finally get the corresponding state value ${V}^{(i)}({o}^{(i)}_t)$ for each agent and handles topological dynamics based on RES. To sum up, we describe the training procedure of \texttt{SVMIX} algorithm in Algorithm 1. \par

\section{Experimental Settings and Numerical Results}
\label{sec:experiment}
In this section, we try to verify the performance of \texttt{SVMIX} in autonomous vehicle control and explain the advantage of our proposed algorithm over other MARL methods. We implement two simulation scenarios on Flow \cite{Flow, Benchmarks}, which is a traffic control benchmarking framework for mixed autonomy traffic. As illustrated in Fig. \ref{fig4}, the ``Figure Eight" scenario and the ``Merge" scenario are chosen as two typical cases to evaluate the performance of our method.

\subsection{The Settings for ``Figure Eight" and Simulation Results}
\label{sec:figure_eight}

\begin{figure}
    \centering
    \subfigcapskip=2pt 
    \subfigure[``Figure Eight"]{
		\includegraphics[width=0.4\linewidth]{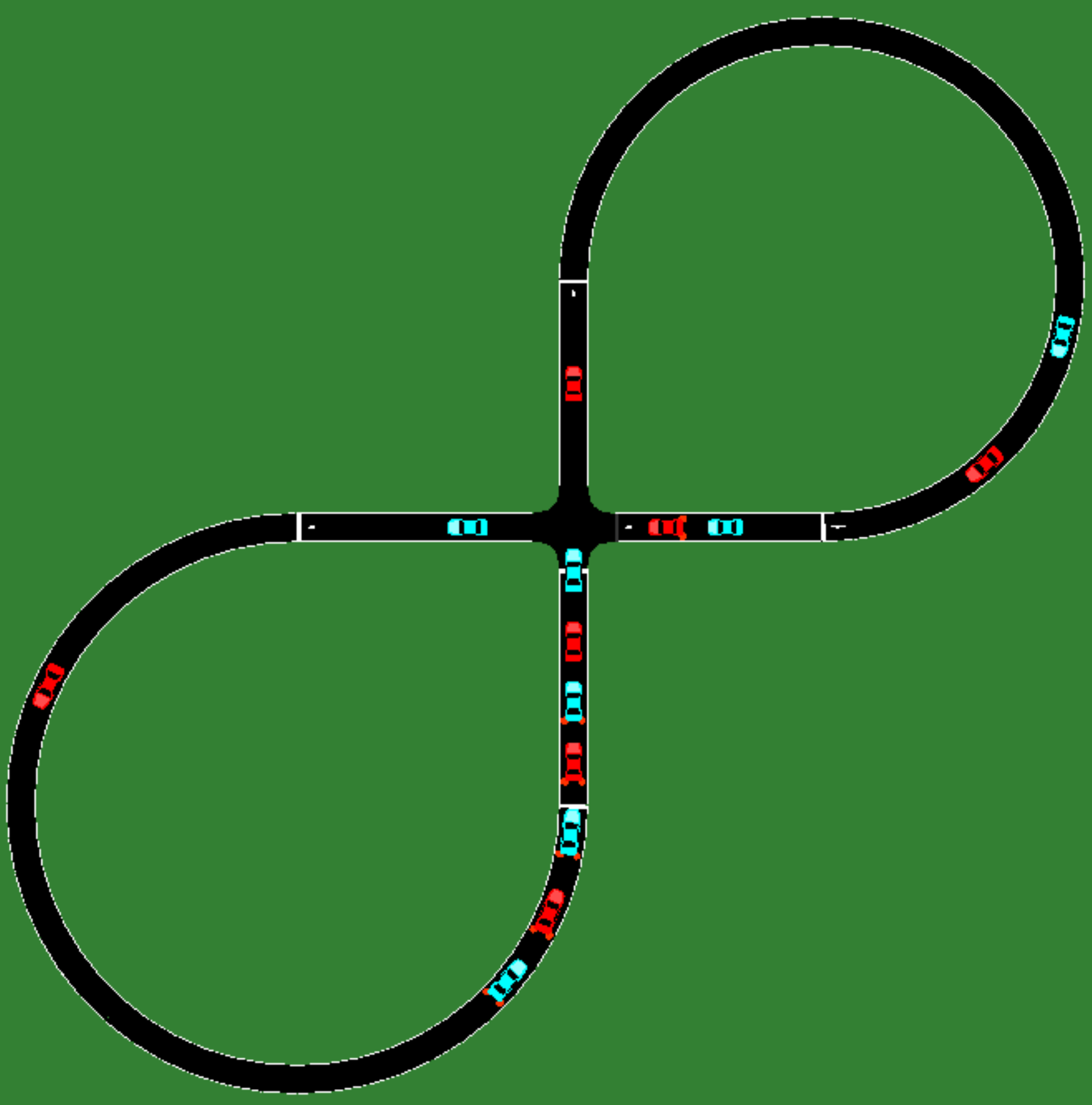}}
  \subfigure[``Merge"]{
		\includegraphics[width=0.55\linewidth]{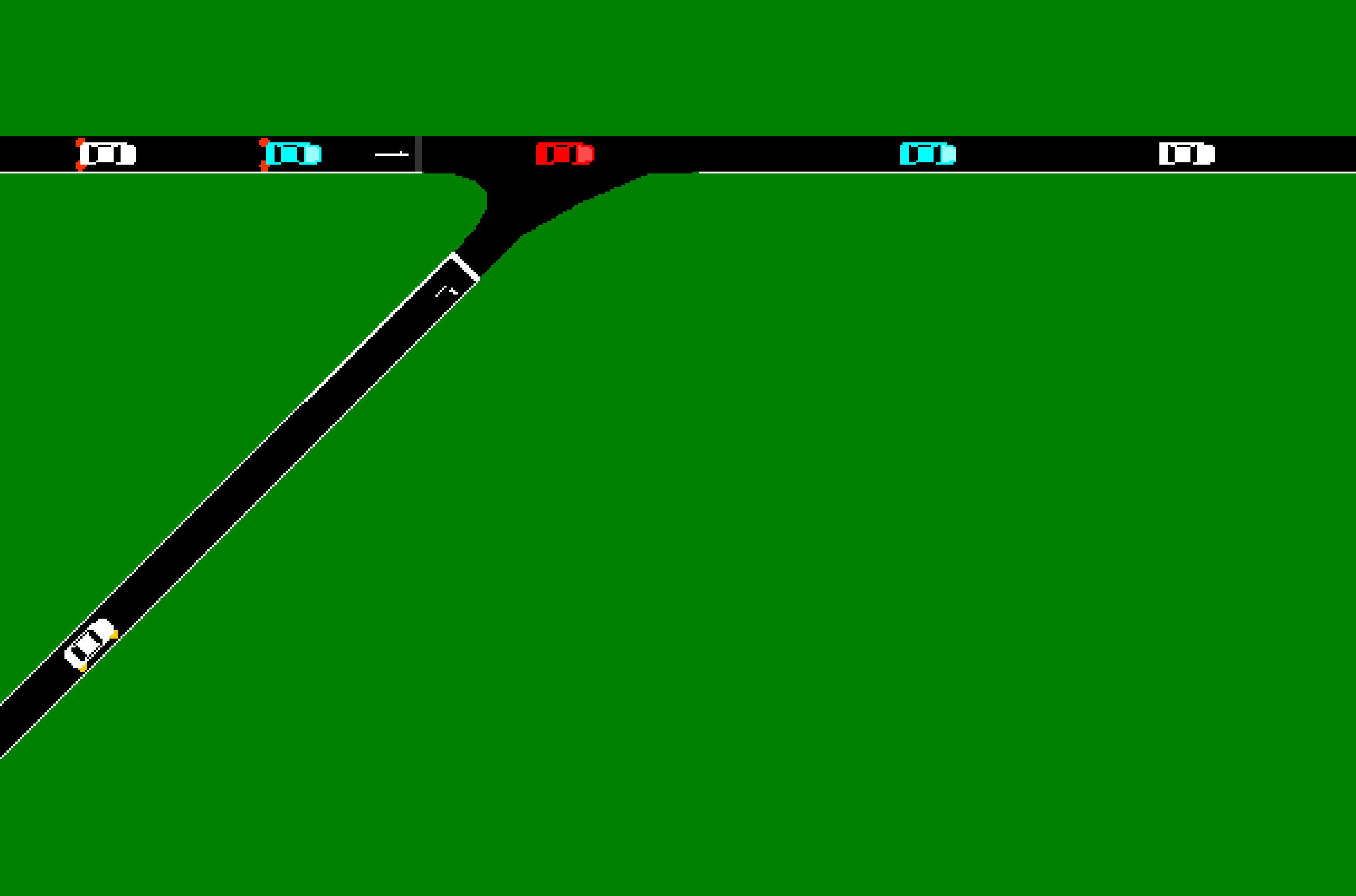}}
            \label{fig4b}
    \caption{The two scenarios for simulations. Here the red vehicles are the DRL-driven vehicles, the blue vehicles are the manned vehicles observed by the DRL-driven vehicles while the white vehicles are the manned vehicles which are not observed in the state space.}
    \label{fig4}
\end{figure}

 The ``Figure Eight" scenario in Fig. \ref{fig4} (a) consists of fixed $M=14$ (i.e., $M_t=14$ for any $t$) vehicles running circularly along a one-way lane with an intersection. So the vehicle must strive for a velocity close to its desired velocity, and timely adjust the velocity when passing through the intersection so as to avoid collisions. What's more, we deploy $7$ manned vehicles and $N=7$ DRL-driven vehicles alternately, where the former vehicles are controlled by Intelligent Driver Model (IDM) defined in \cite{IDM} while the latter ones are controlled by MARL methods. Notably, all the vehicles will perform emergency braking if they are about to crash, and once a collision occurs the episode will be terminated immediately. The corresponding MDP for DRL agents in ``Figure Eight" is defined as below.\par

\paragraph{State and Observation} Here the state contains the information of all the vehicles. Besides, each DRL-driven vehicle can only observe the information of the vehicles ahead and behind. Therefore we have $s_t=\{{v^{(1)}_t,z^{(1)}_t,v^{(2)}_t,z^{(2)}_t, \cdots, v^{(M)}_t,z^{(M)}_t}\}$ and ${o}^{(i)}_t=\{v^{(i_{{\rm ahead}, t})}_t,z^{(i_{{\rm ahead}, t})}_t,v^{(i)}_t,z^{(i)}_t,v^{(i_{{\rm behind}, t})}_t,z^{(i_{{\rm behind}, t})}_t\}$ as the observation of DRL-driven vehicle $i$, where $i_{{\rm ahead}, t}$ and $i_{{\rm behind}, t}$ are the preceding and following vehicles of $i$ at time-step $t$, respectively. \par
\paragraph{Action} As each DRL-driven vehicle only needs to consider the acceleration (i.e. $a^{(i)}=u^{(i)}$), $\boldsymbol{a}_t = \{u^{(1)}, u^{(2)}, \cdots, u^{(N)}\}$. \par
\paragraph{Reward} The reward function is the same as \eqref{eq2} where $\alpha=0$ so the penalty term is ignored. \par
In our setting, the number of episodes $N_{\rm episode}$ is $300$, while each episode has a maximum of $L=1500$ iterations in case of no collision. Also, we design the system utility $U=\frac{1}{L}\sum_{t=1}^{T} r_t$ as a concrete form of (\ref{eq:problem_objectives}) as the evaluation metrics. 
\begin{table}
    \caption{The Key Parameter Settings for ``Figure Eight" and the Algorithms.}
    \label{tb:F8parameters}
    \begin{center}
        \renewcommand\arraystretch{1.5} {
            \begin{tabular}{|c|c|c|}
                \toprule	
                \multicolumn{2}{|c|} {Parameters Definition} & Settings\\
                \midrule	\multicolumn{2}{|c|} {Number of DRL Agents}   & $N=7$ \\
                \hline	\multicolumn{2}{|c|}{Range of Acceleration per Vehicle}   & $\left[-3{\rm m/s^2},3{\rm m/s^2}\right]$                   \\
                \hline	\multicolumn{2}{|c|}{Desired Speed per Vehicle } & $v_d^{(i)} = 20 {\rm m/s}$                    \\
                \hline	\multicolumn{2}{|c|}{Speed Limit per Vehicle} & Up to $30 {\rm m/s}$                    \\
                \hline	\multicolumn{2}{|c|}{Number of Episodes }   & $N_{\rm episode}=300$                   \\
                \hline	\multicolumn{2}{|c|}{Length per Time-step }   & $0.1 {\rm s}$                   \\    
                \hline	\multicolumn{2}{|c|}{Maximal Length per Episode}   & $L=1,500$                   \\
                \hline	\multicolumn{2}{|c|}{Discount Factor}   & $\gamma = 0.99$                   \\
                \hline	\multicolumn{2}{|c|}{Constant in Clip Function}   & $\epsilon = 0.2$                   \\
                \hline	\multicolumn{2}{|c|}{Safety Coefficient in \eqref{eq2}}   & $\alpha = 0$                   \\
                \hline \texttt{FIRL} & Upload Period  & $\tau=10$ \\
                \hline	\multirow{4}{*}{\texttt{SVMIX}} & Number of Filters  & $F=32$                   \\
                \cline{2-3}	& Order per Filter  & $K=3$                   \\
                \cline{2-3}	& Success Probability for RES & $p=0.7$                   \\
                \cline{2-3}	& Number of Parameters per Agent & $N_{\rm para} = 11,653$                   \\
                \bottomrule
            \end{tabular}
        }
    \end{center}
\end{table}

For \texttt{SVMIX}, we use the same architecture as depicted in Fig. \ref{fig2} with $F=32$, $K=3$ and $p=0.7$. Besides, we compare the performance of \texttt{SVMIX} with other MARL methods, including Federated Independent Reinforcement Learning (\texttt{FIRL}) \cite{FIRL}, \texttt{QMIX} \cite{QMIX} and Multi-Graph Attention Network (\texttt{MGAN}) \cite{MGAN}, where \texttt{FIRL} combines federated learning with consensus algorithm based on multiple \texttt{PPO} agents, and \texttt{MGAN} adhibits Graph AttenTion network (GAT) into VD-based MARL architecture. Both \texttt{MGAN} and \texttt{SVMIX} are contingent on a complete graph for inter-agent communication, while \texttt{FIRL} concentrates on a specified graph for consensus. Especially, \texttt{FIRL} uploads the gradient of each agent to a centralized virtual agent every $\tau$ updates. Besides, an optimal baseline is rendered by Flow where all the $14$ vehicles are controlled by IDM, as it is a typical car-following model that contains lots of prior knowledge. It's worth noting that all the methods deal with a global reward, while each agent in \texttt{FIRL} uses the local reward identical as the global reward. Typical parameter settings are summarized in Table \ref{tb:F8parameters}.

\begin{figure}
    \centering
    \subfigbottomskip=2pt 
	\subfigcapskip=-5pt 
    \subfigure[\texttt{FIRL}]{
		\includegraphics[width=0.48\linewidth]{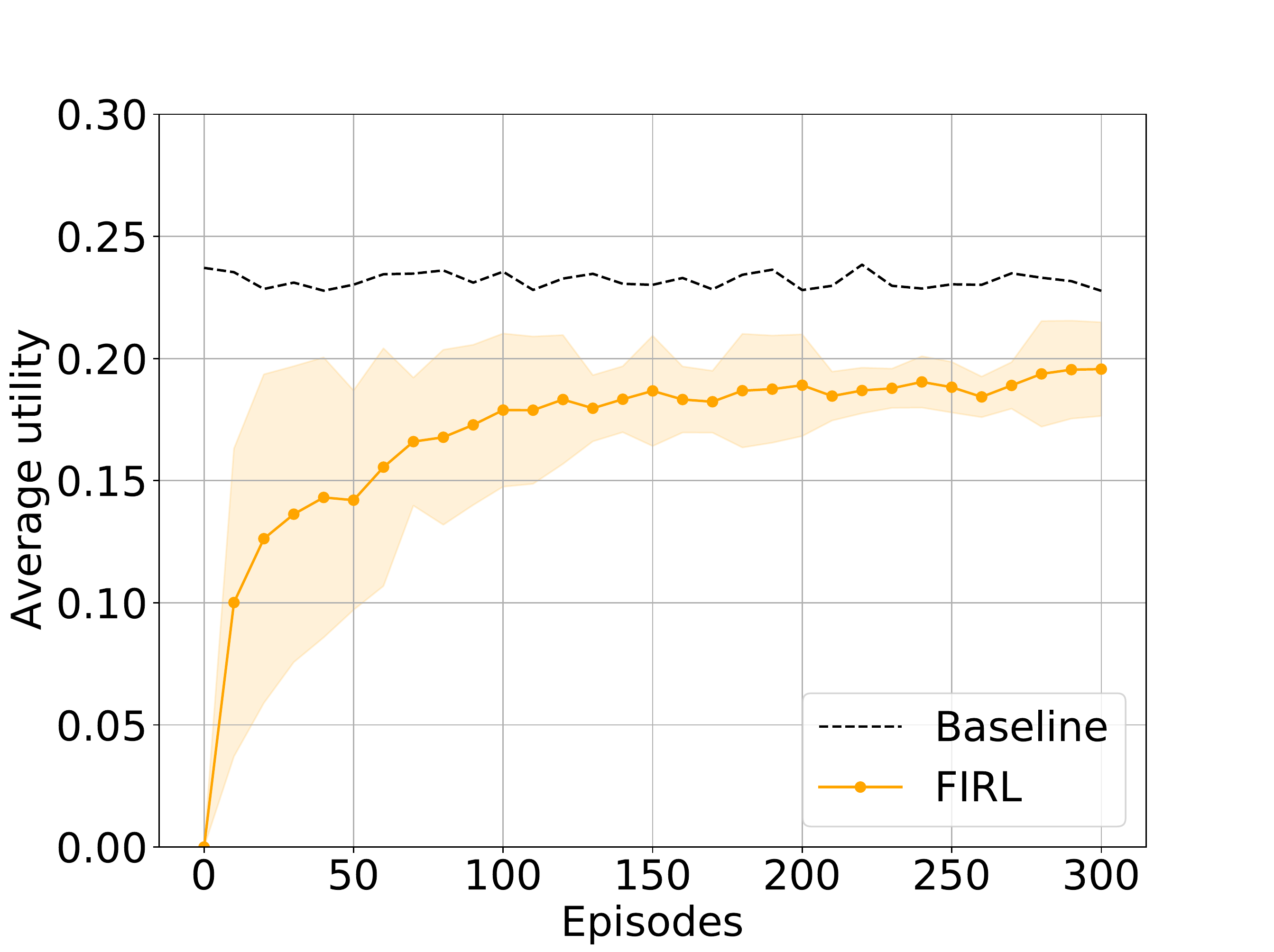}}
  \subfigure[\texttt{QMIX}]{
		\includegraphics[width=0.48\linewidth]{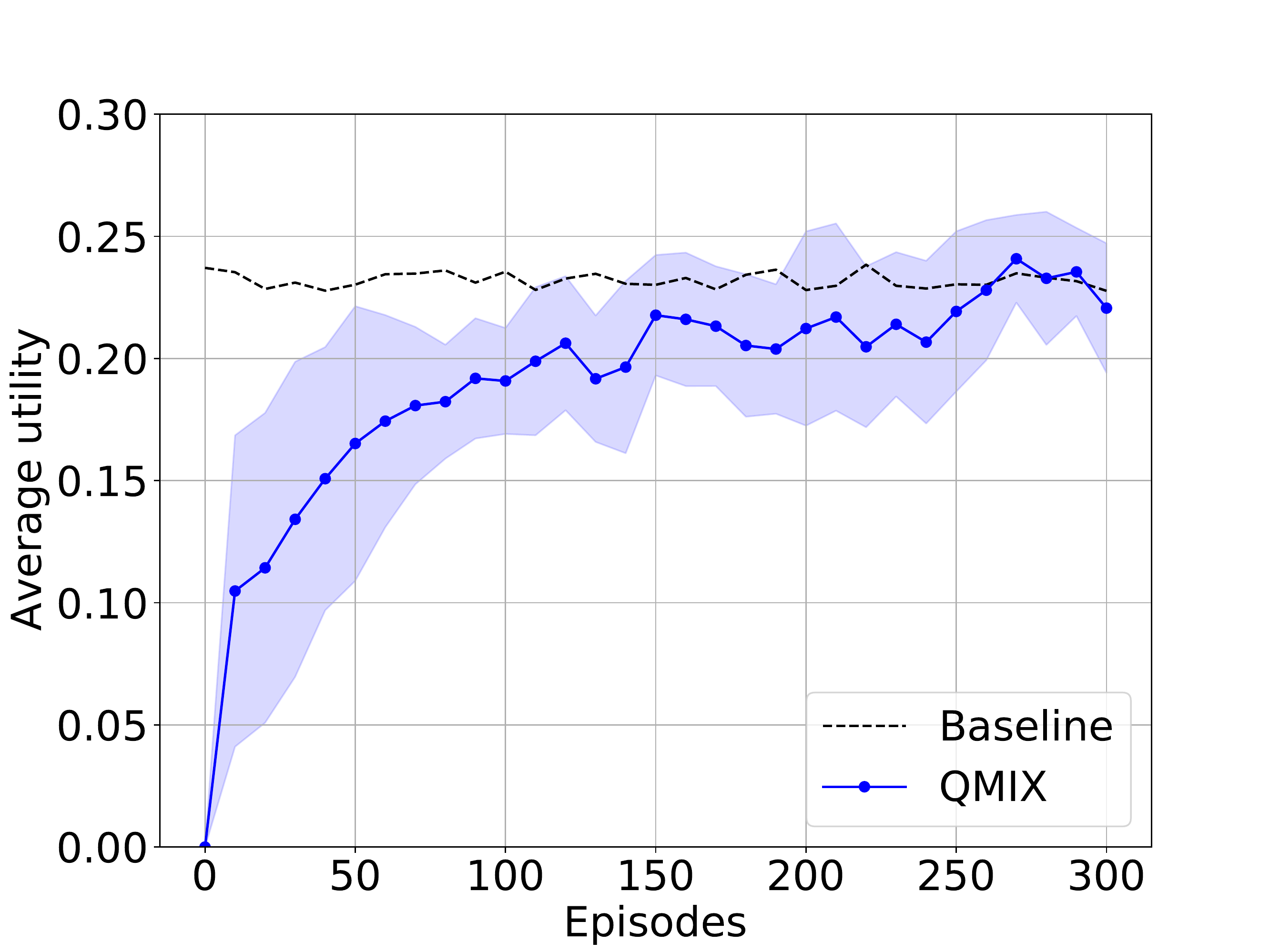}}
  \\
  \subfigure[\texttt{MGAN}]{
		\includegraphics[width=0.48\linewidth]{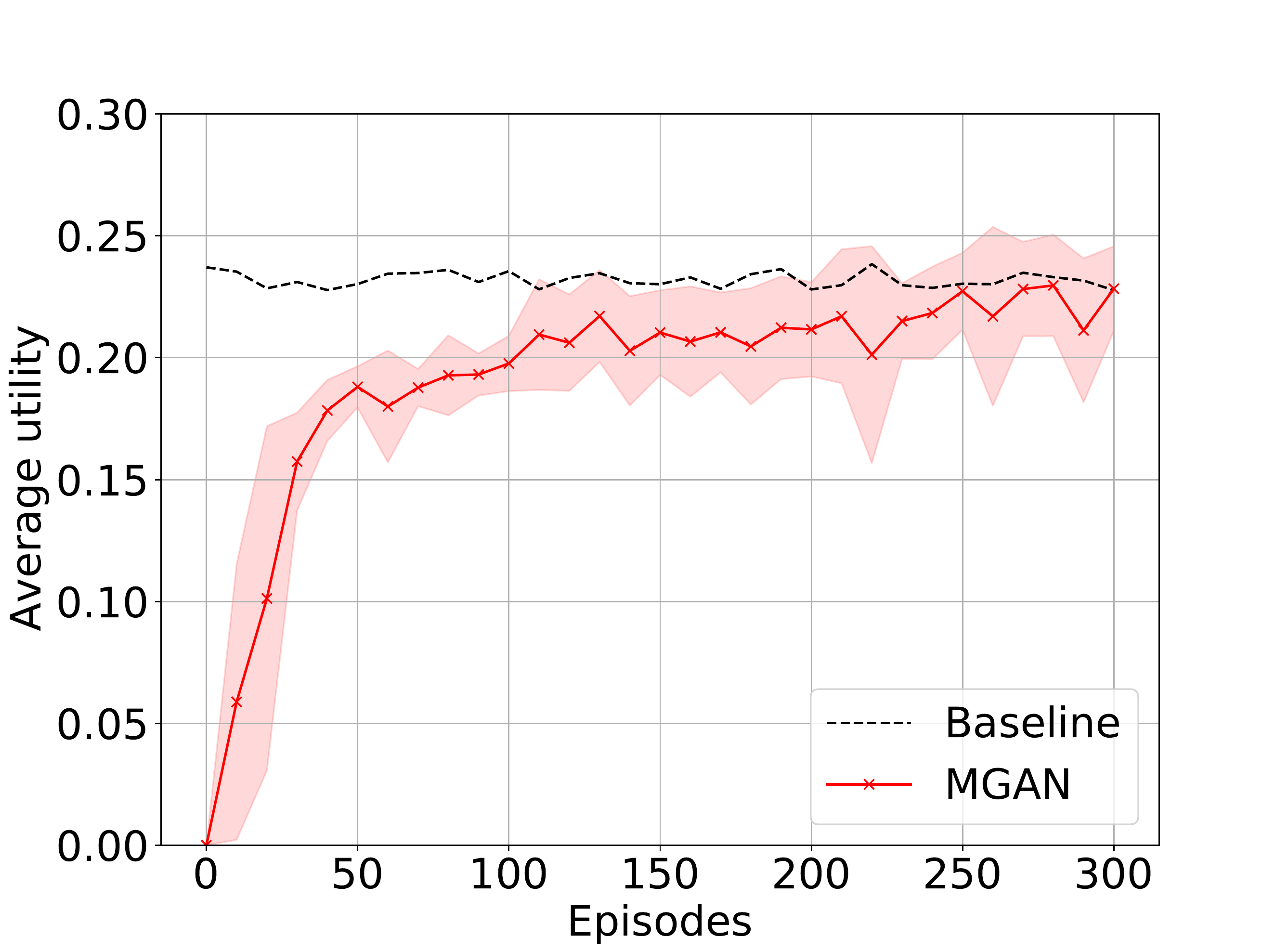}}
  \subfigure[\texttt{SVMIX}]{
		\includegraphics[width=0.48\linewidth]{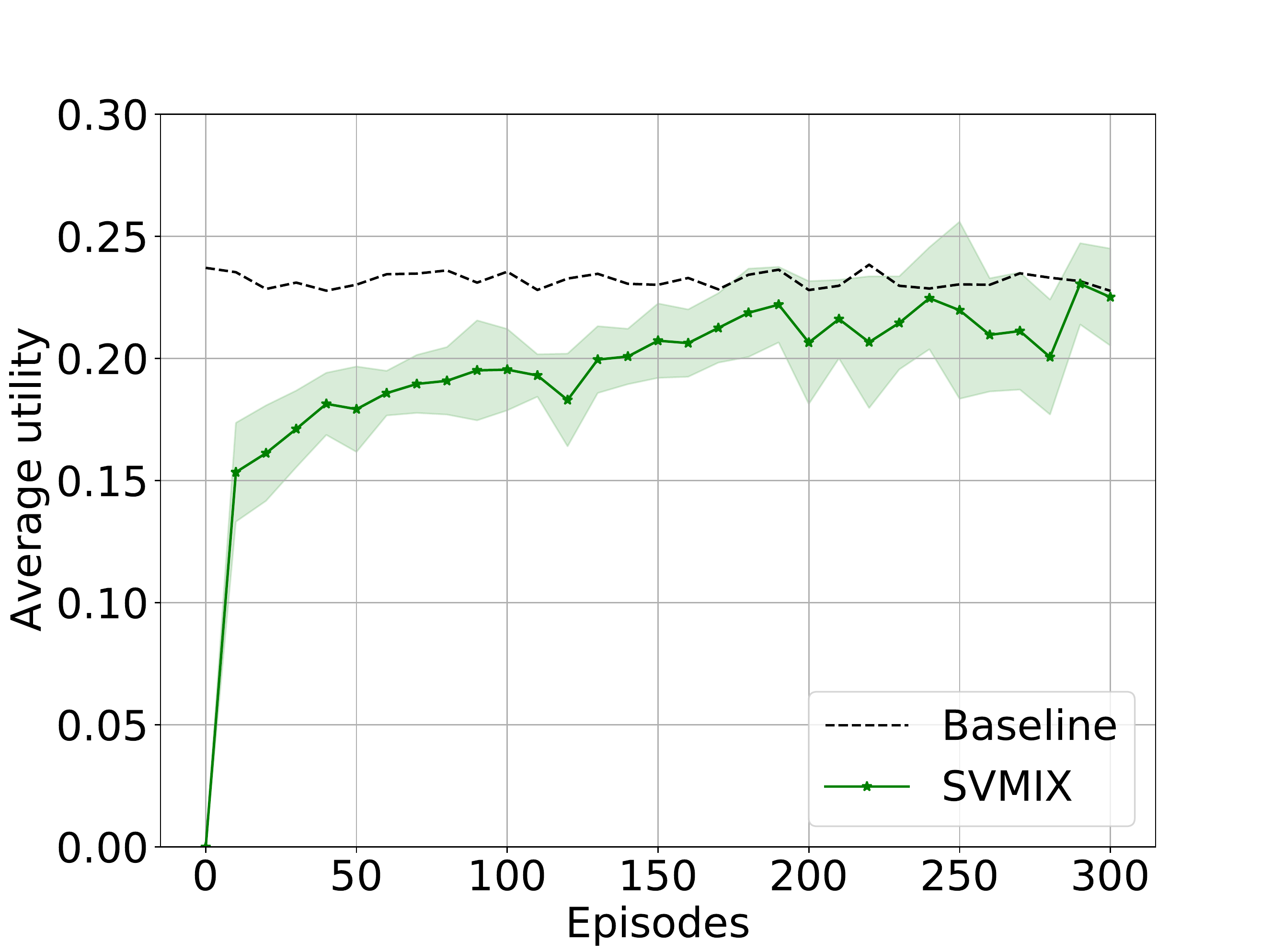}}
    \caption{The average system utility of different methods under the Scenario ``Figure Eight''. Obviously all of the methods converge within about $150$ episodes, where \texttt{FIRL} learns a stable policy at the cost of the convergence speed and utility. \texttt{QMIX} achieves a higher utility thanks to VD and \texttt{MGAN} further improves the convergence speed and reduces the fluctuation owing to the use of GAT. \texttt{SVMIX} outstrips the other methods in terms of the convergence speed and stability with a high utility.}
    \label{fig5}
\end{figure}
Fig. \ref{fig5} compares the average system utility curves of different MARL methods and the optimal baseline in $10$ independent simulations respectively, while Table \ref{tb:mean_and_std} provides the corresponding means and standard deviations of each MARL method. In particular, the utility is computed by averaging the results of another
five testing episodes after every 10 training episodes in a simulation, when each agent directly takes the mean $\mu_i(o^{(i)}_t)$ as a deterministic action. It can be observed from Fig. \ref{fig5} that \texttt{FIRL} yields less utility with the relatively stable performance, while the \texttt{QMIX} achieves a higher utility than \texttt{FIRL}, since the VD learns different contributions of agents with a precise total value function but fluctuates drastically due to the occurrence of collisions. \texttt{MGAN} further improves the convergence speed and somewhat reduces the fluctuations thanks to the captured topological features by GAT. Comparatively, benefiting from the stochastic graph filters with enhanced capability to capture dynamic topological features, \texttt{SVMIX} outperforms the other algorithms as it converges faster with a smaller training variance while learns a safer joint policy with less fluctuations, as well as maintains a high utility.



\begin{table*}
     \caption{The Comparison of Means and Standard Deviations of Utilities in Testing for MARL Methods in ``Figure Eight".}
    \label{tb:mean_and_std}
    \begin{center}
        \renewcommand\arraystretch{1.5} {
            \begin{tabular}{|c|c|c|c|c|c|c|c|}
                \toprule	\multirow{2}{*} {Methods} & \multicolumn{7}{|c|}{The Number of Episodes that have been Experienced for Updating when Testing}       \\
                \cline{2-8} & $30$ & $60$ & $90$ & $120$ & $150$ & $180$ & $210$\\
                \midrule \texttt{FIRL}  & $ 0.136\pm0.060$ & $ 0.156\pm0.049$ & $ 0.173\pm0.033$ & $ 0.183\pm0.026$  & $ 0.187\pm0.023$ & $0.187\pm 0.023$ & $0.185\pm {\bf 0.010}$        \\
                \hline \texttt{QMIX}  & $ 0.134\pm0.065$ & $ 0.174\pm0.043$ & $ 0.192\pm0.025$ & $ {\bf 0.206}\pm0.027$  & $ {\bf 0.218}\pm0.025$ & $0.205\pm 0.029$ & ${\bf 0.217}\pm 0.038$        \\
                \hline	\texttt{MGAN}   & $ 0.157\pm0.020$ & $ 0.180\pm0.023$ & $ 0.193\pm{\bf 0.009}$
 & $ {\bf 0.206}\pm0.020$ & $0.210\pm0.017$ & $0.205\pm 0.024$ & ${\bf 0.217}\pm 0.027$         \\
                \hline	\texttt{SVMIX}   & ${\bf 0.171}\pm{\bf 0.016}$  & ${\bf 0.186}\pm{\bf 0.009}$ & $ {\bf 0.195}\pm0.020$ & $ 0.183\pm{\bf 0.019}$ & $ 0.207\pm{\bf 0.015}$ & ${\bf 0.218}\pm {\bf 0.018}$& $0.216\pm 0.016$              \\
                \bottomrule
            \end{tabular}
        }
    \end{center}
\end{table*}

\begin{table}
    \caption{The Average Communication Overheads of \texttt{FIRL} and \texttt{SVMIX}.}
    \label{tb:communicationOverhead}
    \begin{center}
        \renewcommand\arraystretch{1.5} {
            \begin{tabular}{|c|c|c|c|}
                \toprule	Methods & Communication Overheads                                     \\
                \midrule	\texttt{FIRL}    & $\frac{1}{\tau}\left(N*N_{\rm para}\right) =  8,157.1$ \\
                \hline	\texttt{SVMIX}   & $ N * N_{\rm epoch}* N_{\rm batch} = 7,168$                   \\
                \bottomrule
            \end{tabular}
        }
    \end{center}
\end{table}
In terms of the communication overhead, Table \ref{tb:communicationOverhead} compares \texttt{SVMIX} and \texttt{FIRL}, where both of the methods will be updated once with a batch containing $N_{\rm batch}$ samples. Notably, the VD-based methods have almost the same communication overheads.
Consequently, \texttt{SVMIX} reduces $12.13\%$ of the communication overheads compared to \texttt{FIRL} as it only needs to transmit a batch of values rather than the gradients of all the parameters. The reason lies in that the communication overheads of \texttt{FIRL} are proportional to the number of parameters (i.e., $N_{\rm para}$) while the communication overheads of \texttt{SVMIX} are proportional to the number of batches (i.e., $N_{\rm batch}$) and the number of epochs (i.e., $N_{\rm epoch}$). 

\subsection{The Settings for ``Merge" and Simulation Results}
\label{sec:merge}
As depicted in Fig. \ref{fig4} (b), the ``Merge" scenario simulates the on-ramp merging on the highway. Similar to ``Figure Eight", it's essential for vehicles to consider how to avoid collisions and congestion due to the meeting at the merging point. As ``Merge" is an unclosed network, there assumes to exist a traffic flow where the vehicles frequently flow in and out, leading to the variations in the number of vehicles. Following the definition of the model in Section \ref{sec:background}, the scenario is set to allow at most $2,100$ vehicles to flow in per hour, including $2,000$ vehicles at most on the trunk road and $100$ vehicles at most on the ramp per hour. In particular, among the $2,000$ vehicles, $25\%$ of them will be assigned as DRL-driven vehicles. Consistent with the MDP defined in Section \ref{sec:background}, we consider an MDP as follows.

\begin{table}
    \caption{The Key Parameter Settings for ``Merge" and the Algorithms.}
    \label{tb:Mergeparameters}
    \begin{center}
        \renewcommand\arraystretch{1.5} {
            \begin{tabular}{|c|c|c|}
                \toprule	
                \multicolumn{2}{|c|} {Parameters Definition} & Settings\\
                \midrule	\multicolumn{2}{|c|} {Number of DRL Agents}   & $N=13$ \\
                \hline	\multicolumn{2}{|c|}{Maximum of Vehicles per Hour}   & $2100$                   \\
                \hline	\multicolumn{2}{|c|}{Proportion of DRL-driven Vehicles}   & $ 25\% $                   \\
                \hline	\multicolumn{2}{|c|}{Range of Acceleration per Vehicle}   & $\left[-1.5{\rm m/s^2},1.5{\rm m/s^2}\right]$                   \\
                \hline	\multicolumn{2}{|c|}{Desired Speed per Vehicle } & $v_d^{(i)} = 20 {\rm m/s}$                    \\
                \hline	\multicolumn{2}{|c|}{Speed Limit per Vehicle} & Up to $30 {\rm m/s}$                    \\
                \hline	\multicolumn{2}{|c|}{Number of Episodes }   & $N_{\rm episode}=300$                   \\
                \hline	\multicolumn{2}{|c|}{Length of per Time-step }   & $1 {\rm s}$                   \\                
                \hline	\multicolumn{2}{|c|}{Maximal Length per Episode}   & $L=750$                   \\
                \hline	\multicolumn{2}{|c|}{Discount Factor}   & $\gamma = 0.99$                   \\
                \hline	\multicolumn{2}{|c|}{Constant in Clip Function}   & $\epsilon = 0.2$                   \\
                \hline	\multicolumn{2}{|c|}{Safety Coefficient in \eqref{eq2}}   & $\alpha = 0.1$                   \\
                \hline \texttt{FIRL} & Upload Period  & $\tau=10$ \\
                \hline	\multirow{4}{*}{\texttt{SVMIX}} & Number of Filters  & $F=32$                   \\
                \cline{2-3}	& Order per Filter  & $K=3$                   \\
                \cline{2-3}	& Success Probability for RES& $p=0.7$                   \\
                \bottomrule
            \end{tabular}
        }
    \end{center}
\end{table}

\paragraph{State and Observation} Here the state contains the information of the observable vehicles (just like the red and blue vehicles in Fig. \ref{fig4} (b)) instead of all the vehicles. Also,  each DRL-driven vehicle can only observe the information of the preceding and following vehicles. So $s_t$ only consists of the positions and velocities of DRL-driven vehicles as well as the vehicles ahead of and behind them. Furthermore, we fix the number of algorithmically involved DRL-driven vehicles as $N=13$ and if the practical number of vehicles $N_t > N$, the other $N_t - N$ vehicles will be treated as manned vehicles; otherwise, the state will be padded with zeros, as if there are $N$ DRL-driven vehicles.
The dimension of the state space consequently remain unchanged. Therefore the state is represented by $s_t=\{{o}^{(1)}_t, {o}^{(2)}_t, \cdots, {o}^{(N)}_t\}$ where ${o}^{(i)}_t=\{v^{(i_{{\rm ahead}, t})}_t,z^{(i_{{\rm ahead}, t})}_t,v^{(i)}_t,z^{(i)}_t,v^{(i_{{\rm behind}, t})}_t,z^{(i_{{\rm behind}, t})}_t\}$ expresses the observation of DRL-driven vehicle $i$ where $i=1, 2, \cdots, N$. \par
\paragraph{Action} Same as the above, the $N=13$ DRL-driven vehicles will choose their actions according to their policies so the dimension of the joint action space is also immutable and $\boldsymbol{a}_t = \{u^{(1)}, u^{(2)}, \cdots, u^{(N)}\}$. When $N_t > N$, the first $N$ vehicles will be treated as DRL-driven vehicles and the others are regarded as manned vehicles. In case $N_t < N$, the extra actions will be ignored during the interaction with the environment. It's worth noting that the underlying graph $\mathcal{G}^+$ is still a complete graph with fixed $N$ vertices.\par
\par
\paragraph{Reward} The reward function is the same as \eqref{eq2} where $\alpha=0.1$ and the penalty term is defined as
\begin{equation}
    \label{Penalty}
    D(z^{(1)}, \cdots, z^{(M_t)})=\sum_{j=1}^{M_t}\operatorname{max}[C_2 -\Vert z^{(i_{\rm {ahead}, t})} - z^{(i)}\Vert_2, 0],
\end{equation}
where $C_2$ is a constant which represents the desired following distance of each vehicle. \par
In our setting, the number of episodes $N_{\rm episode}$ is 300, while each episode has a maximum of $L=750$ iterations in case of no collision. 
For \texttt{SVMIX}, we use the same architecture as in Fig. \ref{fig2} with $F=32$, $K=3$ and $p=0.7$. Besides, we compare the performance of \texttt{SVMIX} with \texttt{FIRL}, \texttt{QMIX} and \texttt{MGAN} with the same detailed setting described earlier in Section \ref{sec:figure_eight}. Typical parameter settings are summarized in Table \ref{tb:Mergeparameters}.

\begin{figure}
    \centering
    \subfigbottomskip=2pt 
	\subfigcapskip=-5pt 
    \subfigure[\texttt{FIRL}]{
		\includegraphics[width=0.48\linewidth]{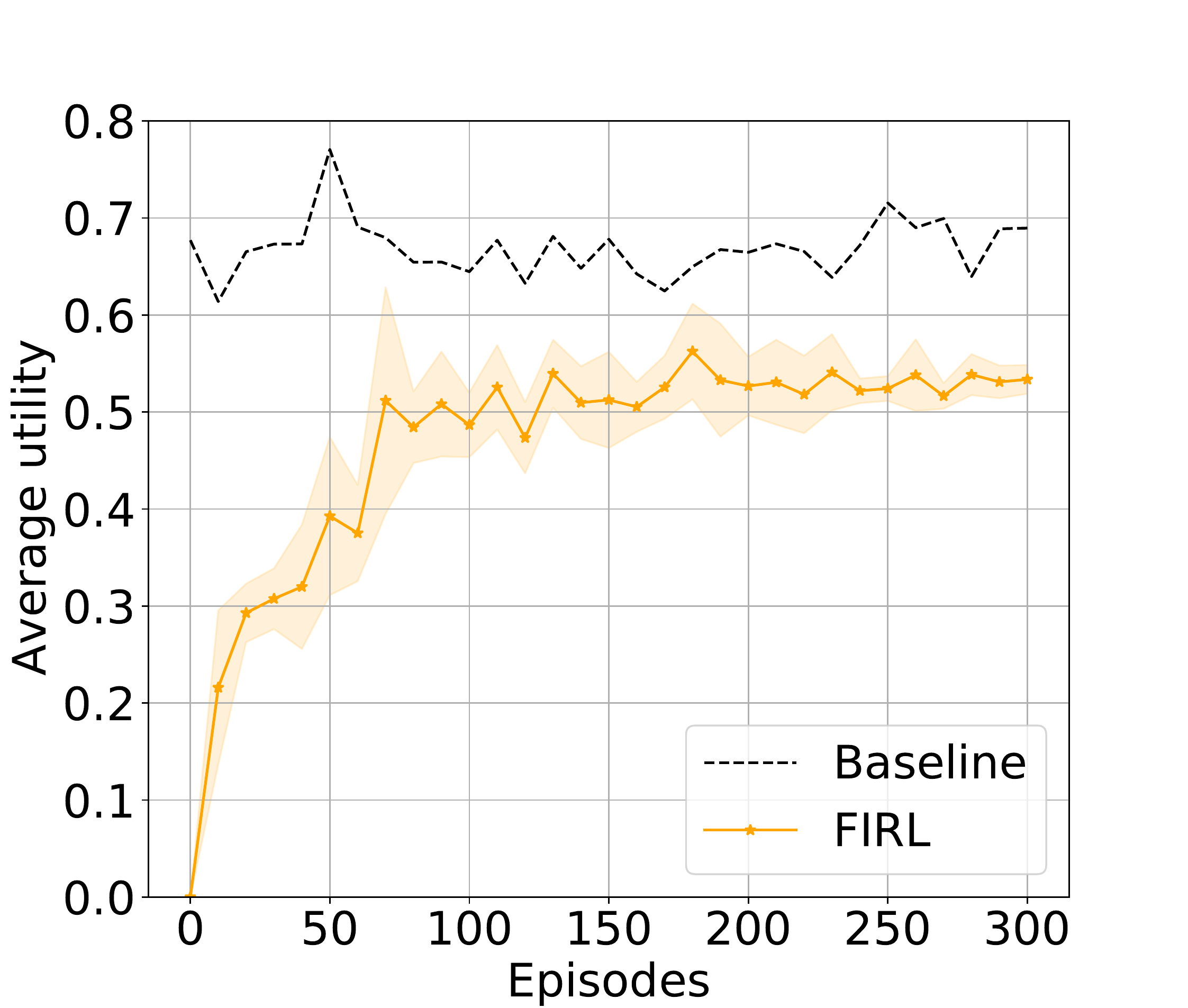}}
  \subfigure[\texttt{QMIX}]{
		\includegraphics[width=0.48\linewidth]{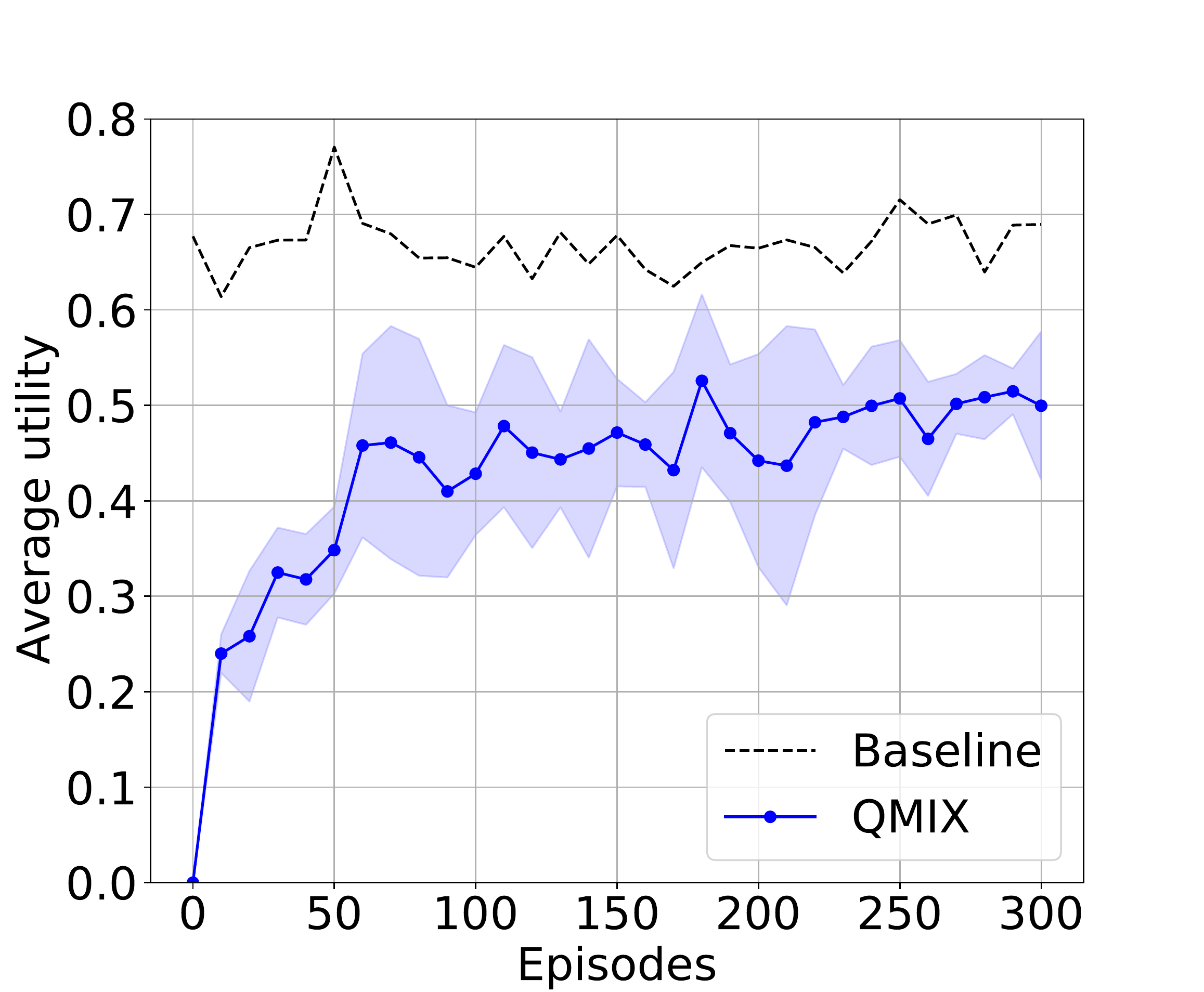}}
  \\
  \subfigure[\texttt{MGAN}]{
		\includegraphics[width=0.48\linewidth]{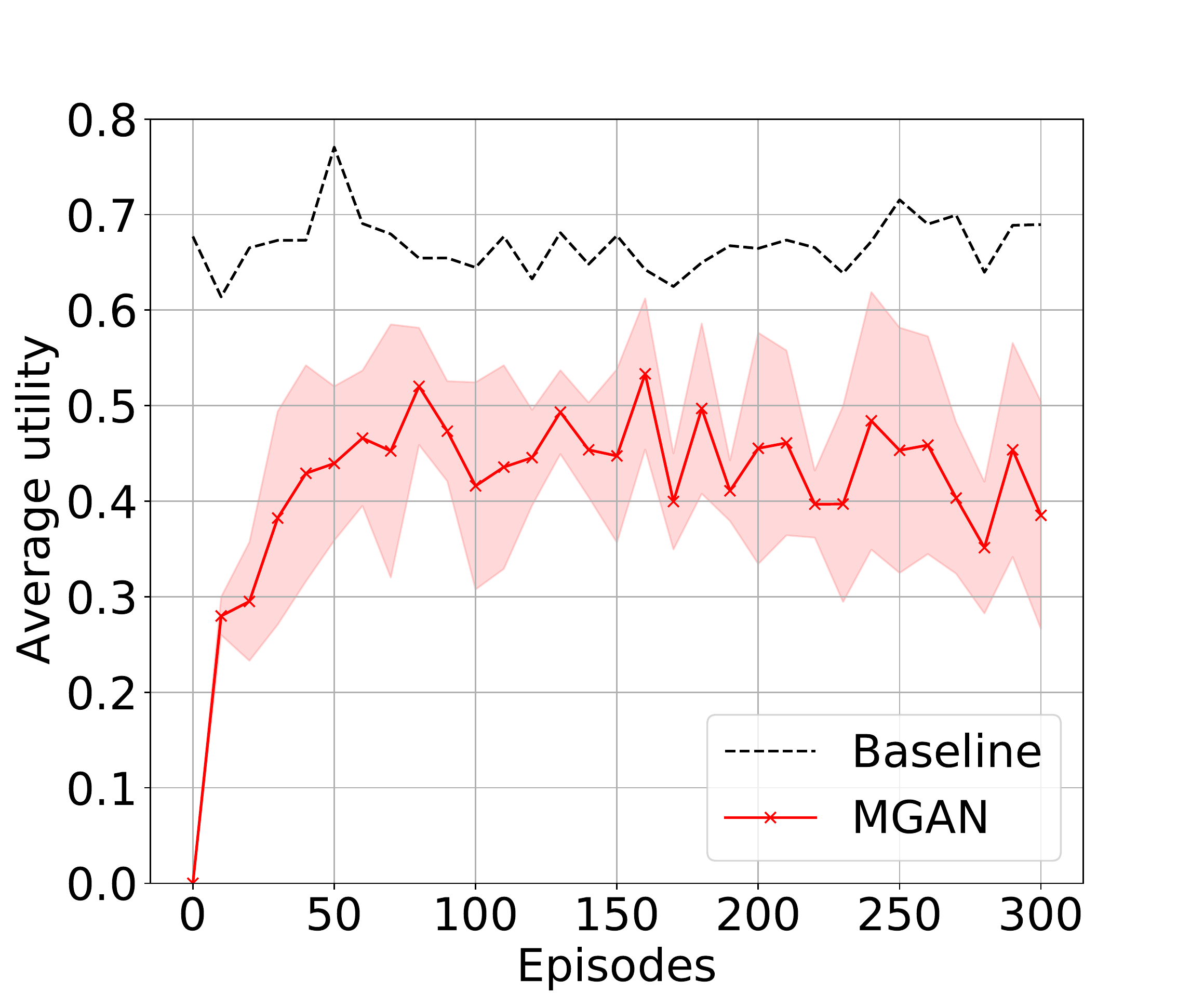}}
  \subfigure[\texttt{SVMIX}]{
		\includegraphics[width=0.48\linewidth]{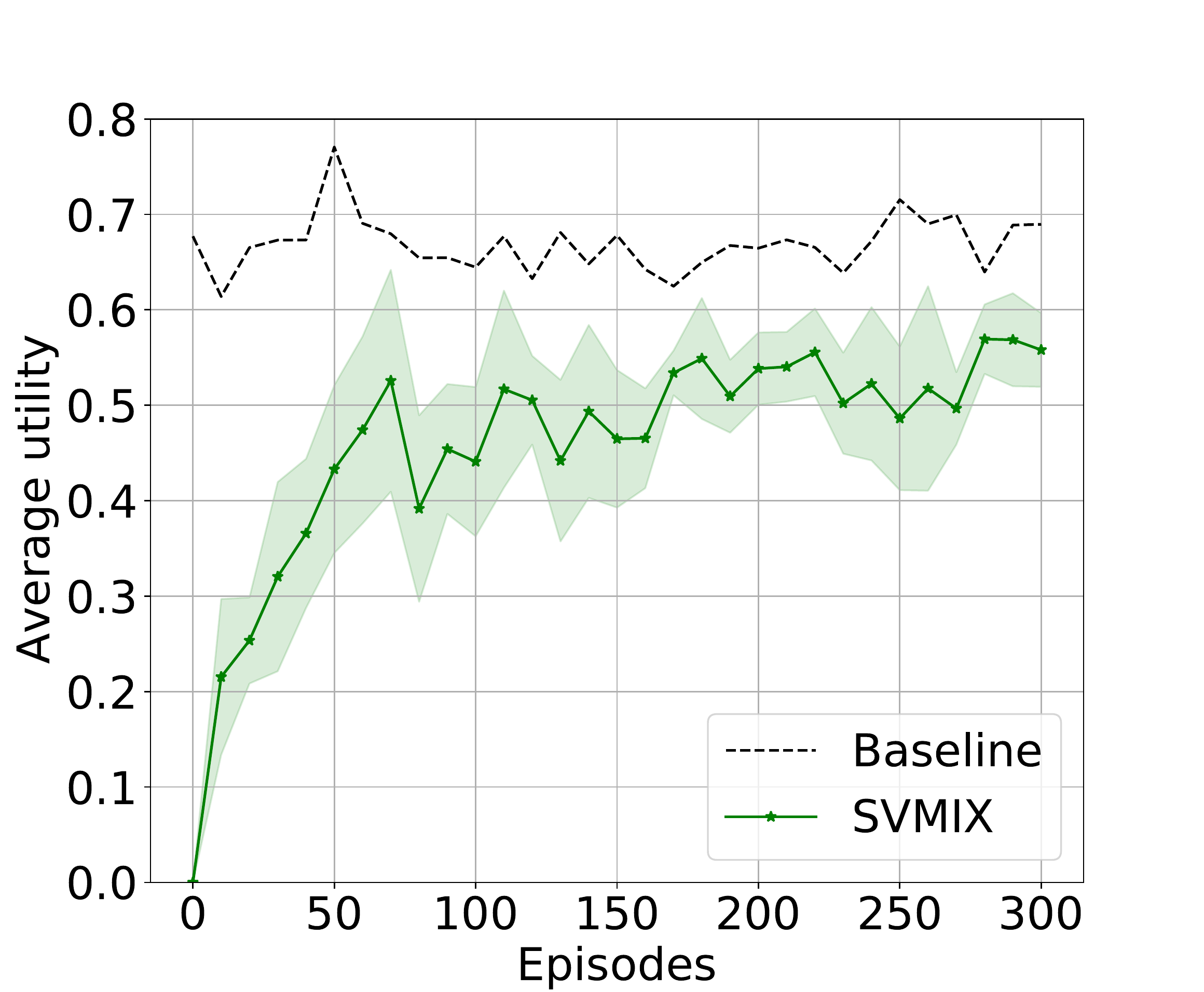}}
    \caption{The average system utility of different methods under the Scenario ``Merge''. All the methods converge within about $100$ episodes. \texttt{FIRL} still learns a stable policy and reaches a not bad average utility. \texttt{MGAN} converges quickly but seems unstable in terms of the undulant and low average utility.  \texttt{QMIX} and \texttt{SVMIX} both hold a more stable performance, where the average utility of \texttt{SVMIX} is slightly better than \texttt{QMIX}.}
    \label{fig6}
\end{figure}

\begin{table*}
     \caption{The Comparison of Means and Standard Deviations of Utilities in Testing for MARL Methods in ``Merge".}
    \label{tb:mean_and_std_merge}
    \begin{center}
        \renewcommand\arraystretch{1.5} {
            \begin{tabular}{|c|c|c|c|c|c|c|c|}
                \toprule	\multirow{2}{*} {Methods} & \multicolumn{7}{|c|}{The Number of Episodes that have been Experienced for Updating when Testing}       \\
                \cline{2-8} & $30$ & $60$ & $90$ & $120$ & $150$ & $180$ & $210$\\
                \midrule \texttt{FIRL}  & $ 0.308\pm{\bf 0.031}$ 
                & $ 0.375\pm{\bf 0.050}$ 
                & $ {\bf 0.508}\pm 0.054$ 
                & $ 0.473\pm{\bf 0.037}$ 
                & $ {\bf 0.512}\pm 0.050$ 
                & ${\bf 0.562}\pm {\bf 0.050}$       
                & $0.531\pm 0.044$\\
                \hline	\texttt{QMIX}  & $ 0.325\pm0.047$ 
                & $ 0.458\pm0.096$ 
                & $ 0.410\pm 0.090$ 
                & $ 0.450\pm0.100$ 
                & $ 0.471\pm 0.056$ 
                & $0.526\pm 0.091$       
                & $0.437\pm 0.146$\\
                \hline	\texttt{MGAN}   & $ {\bf 0.382}\pm0.112$ 
                & $ 0.466\pm0.071$ 
                & $ 0.473\pm {\bf 0.052}$
                & $0.446\pm0.046$ 
                & $0.447\pm {\bf 0.009}$ 
                & $0.497\pm 0.089$  
                & $0.461\pm 0.097 $\\
                \hline	\texttt{SVMIX}   & $ 0.320\pm 0.099$  
                & $ {\bf 0.474}\pm 0.098$ 
                & $ 0.454\pm 0.068$ 
                & $ {\bf 0.505}\pm0.051$ 
                & $ 0.465\pm 0.072$
                & $ 0.549\pm 0.063$     
                & $ {\bf 0.540} \pm {\bf 0.037}$\\
                \bottomrule
            \end{tabular}
        }
    \end{center}
\end{table*}
Fig. \ref{fig6} compares the average system utility curves of different MARL methods and the optimal baseline in $5$ independent simulations respectively, while Table \ref{tb:mean_and_std_merge} provides the corresponding means and standard deviations of each MARL method. In particular, the utility is computed by averaging the results of another
five testing episodes after every $10$ training episodes in a simulation, when each agent directly takes the mean $\mu_i(o^{(i)}_t)$ as a deterministic action. Notably, \texttt{MGAN} achieves an unstable and the worst performance shown in Fig. \ref{fig6}(b), and even worse than \texttt{QMIX} as well as \texttt{FIRL}. Especially at the last $150$ episodes, the curve of \texttt{MGAN} fluctuates violently. This is possibly because the DRL-driven vehicles are irregularly allocated due to the traffic flow and it becomes a degenerative feedback to the learning of attention coefficients. Both \texttt{QMIX} and \texttt{SVMIX} generates a relatively stable curve with higher average utilities than \texttt{FIRL} and \texttt{MGAN}. What's more, benefiting from the stochastic graph filters, \texttt{SVMIX} outperforms \texttt{QMIX} with higher utility and lower variance according to Table \ref{tb:mean_and_std_merge}. It is noteworthy that the performance of \texttt{SVMIX} obviously outperforms \texttt{MGAN} in ``Merge", which verifies the aforementioned idea that it's necessary to deal with the dynamic topology if we want to capture topological features. Compared with \texttt{FIRL}, \texttt{SVMIX} has similar means of average utilities and seems more unstable as the variances are larger in the scenario. More meaningfully, \texttt{SVMIX} still superiors in terms of communication overheads. 

\subsection{Hyperparameters Adjusting for \texttt{SGNN} in \texttt{SVMIX}}
\label{sec:Hyper}
\begin{figure*}
    \centering
    \subfigbottomskip=2pt 
	\subfigcapskip=-15pt 
    \subfigure[Probability $p$]{
		\includegraphics[width=0.32\linewidth]{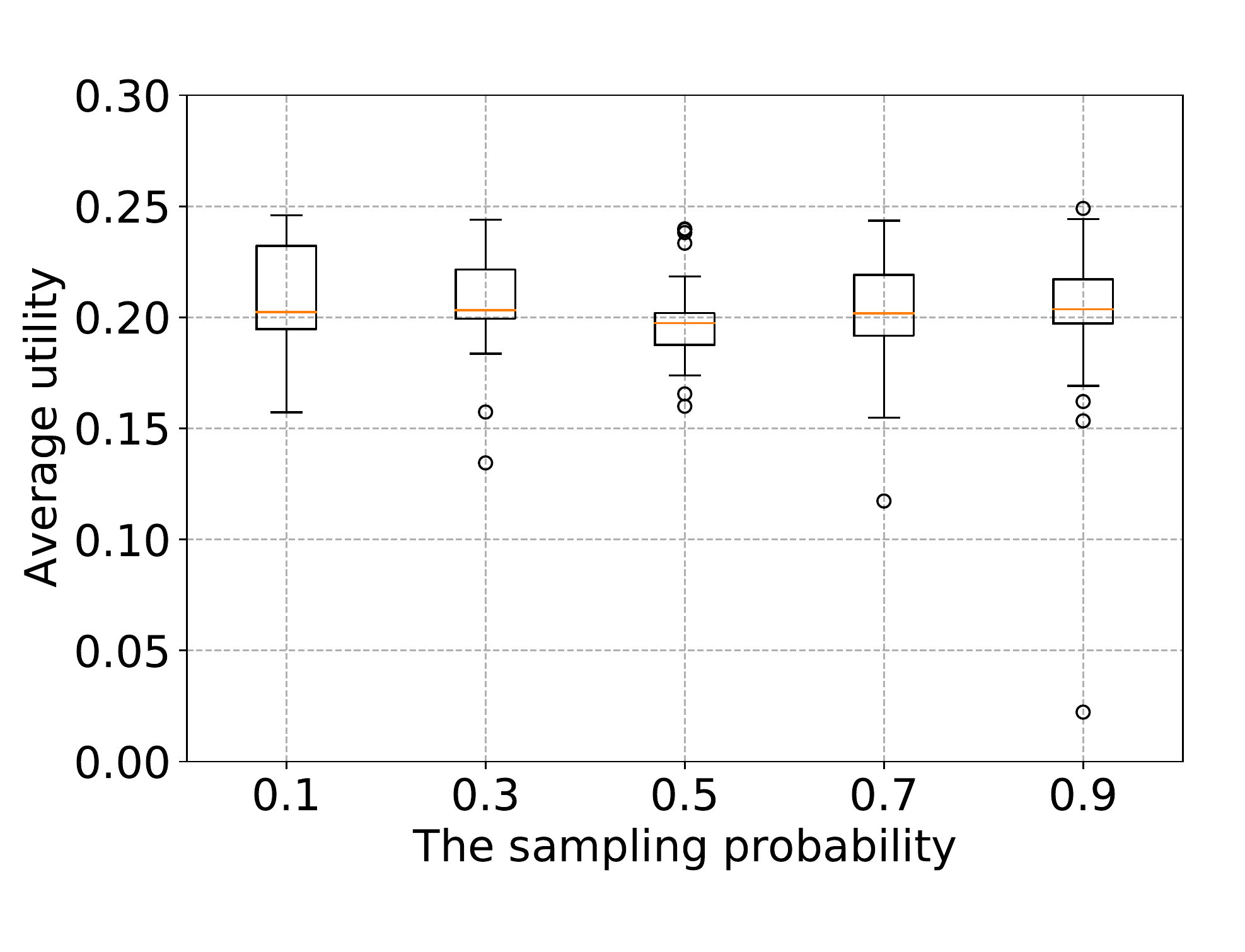}}
  \subfigure[Order of the filter $K$]{
		\includegraphics[width=0.32\linewidth]{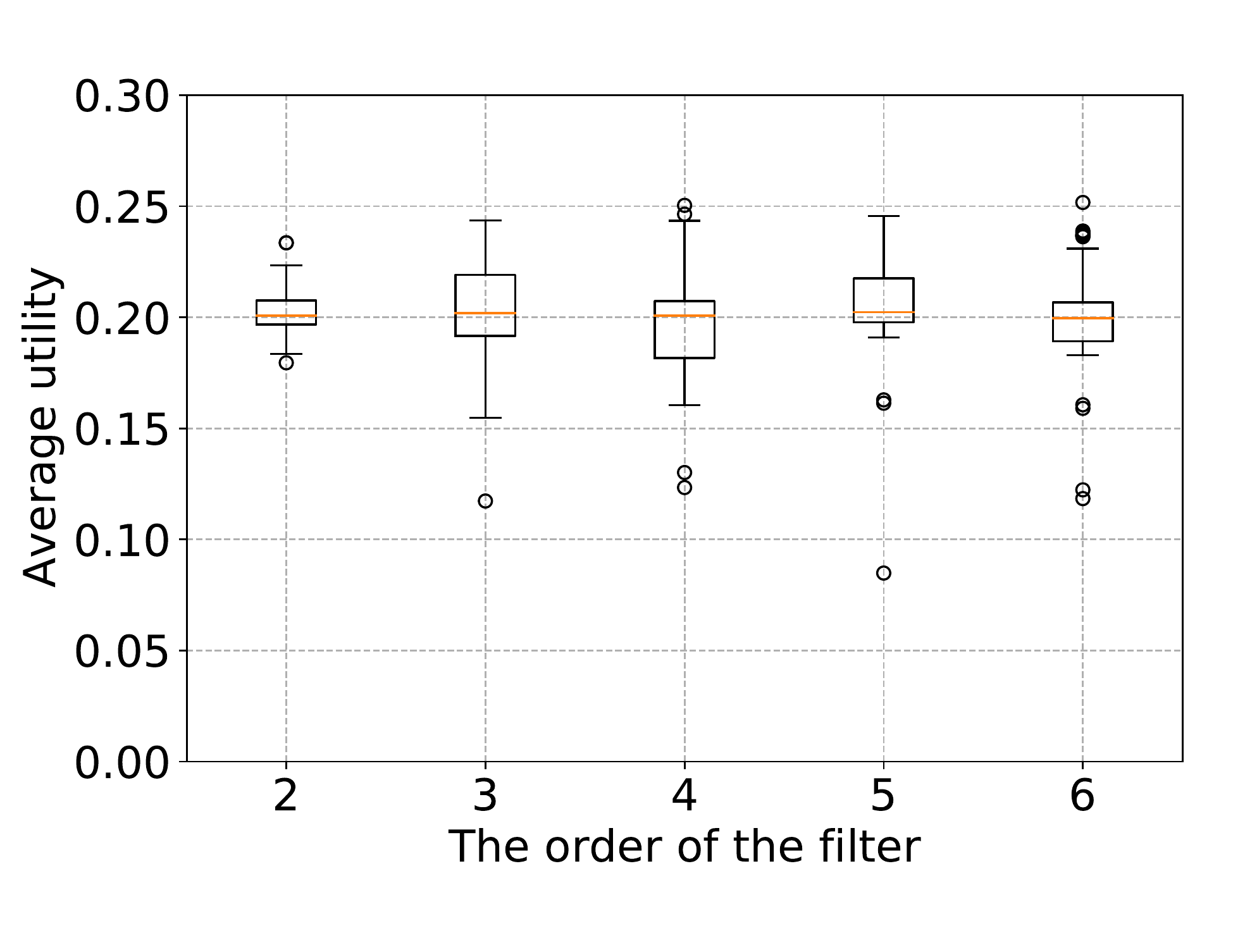}}
  \subfigure[Number of filters $F$]{
		\includegraphics[width=0.32\linewidth]{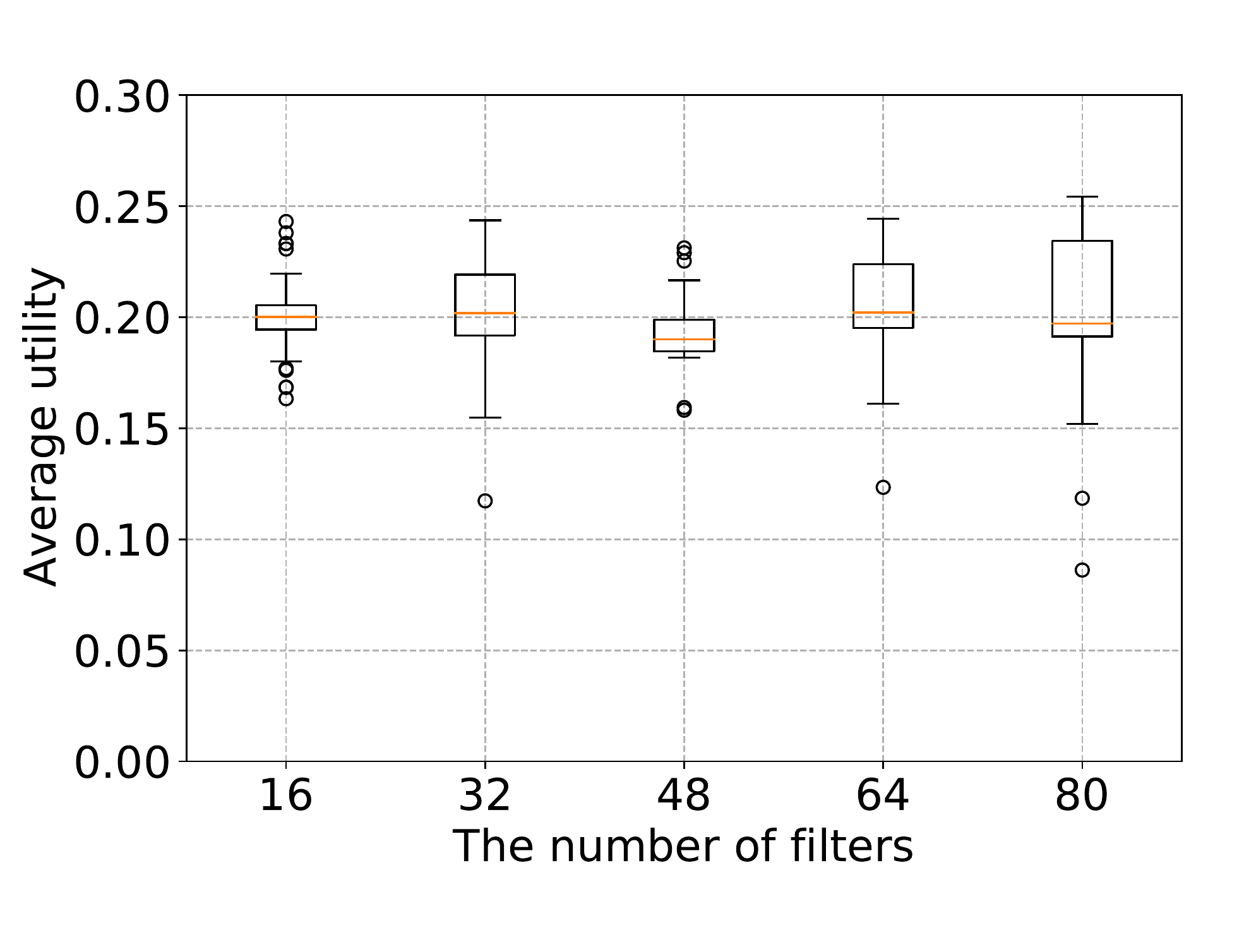}}
    \caption{The average utility of \texttt{SVMIX} with different structure of \texttt{SGNN} under the corresponding hyperparameters by changing (a) the sampling probability $p$, (b) the order of each filter $K$ and (c) the number of filters $F$ from the testing episodes after $100, 110, \cdots, 200$ training episodes through $5$ independent simulations.}
    \label{fig7}
\end{figure*}

To clarify the influence of the hyperparameters in \texttt{SGNN}, we further carry out supplementary experiments in ``Figure Eight" by changing one hyperparameter and keeping the others the identical. Fig. \ref{fig7} provides the corresponding results where we get the results from the testing episodes after $100, 110, \cdots, 200$ training episodes (i.e., when \texttt{SVMIX} converges slowly) respectively through $5$ independent simulations. 

\paragraph{Probability $p$} As shown in Fig. \ref{fig7}(a), keeping $K=3$ and $F=32$, we evaluate the performance of \texttt{SVMIX} for $p\in\{0.1, 0.3, 0.5, 0.7, 0.9\}$. For a Bernoulli-sampled RES, the variance of the outputs from \texttt{SGNN} is proportional to $p(1-p)$ and will be maximized when $p=0.5$. Obviously, the result of $p=0.5$ is the most unstable as there are many outliers. The settings of $p=0.3$ and $p=0.7$ yield the similar stable performance, while the performance of $p=0.1$ and $p=0.9$ seem more unstable. This is because when $p=0.1$ and $p=0.9$, \texttt{SGNN} is unable to make sufficient exploration due to the too small or too large probability $p$ based on RES. Thus a probability like $p=0.7$ sounds more appropriate for \texttt{SGNN}.
\paragraph{Order of the filter $K$} Fig. \ref{fig7}(b) demonstrates the influence of the filter order $K$ by retaining $p=0.7$ and $F=32$ with $K$ spanning from $2$ to $6$. Consistent with the analysis in \cite{SGNN}, which states that the variance of the outputs from \texttt{SGNN} is proportional to $K$, the variance of the system utility also becomes greater as $K$ grows. 
Thus an order like $K=2$ or $K=3$ will be suitable for ``Figure Eight".
\paragraph{Number of filters $F$} Fig. \ref{fig7}(c) demonstrates the influence of the number of filters $F$ by retaining $p=0.7$ and $K=3$ with $F\in\{16, 32, 48, 64, 80\}$. For $F=16$, the capability of \texttt{SGNN} for feature capturing decreases and thus leads to an unstable performance. As $F$ increases, \texttt{SVMIX} generally gives better performance, though it slightly suffers from the deficiency of training samples to train a larger neural network.
\section{Conclusion and Discussion}
\label{sec:conclusion}
This paper aims to address reward assignment problem in a dynamic environment by Value Decomposition and tentatively puts forward an \texttt{SGNN} based multi-agent actor-critic architecture called \texttt{SVMIX} with \texttt{PPO} as the individual agent. In particular, \texttt{SGNN}, which consists of parallel stochastic graph filters, has been leveraged to enhance the resilience to the environment volatility and thus capably capture dynamic underlying features in a more effective manner. To demonstrate the feasibility of \texttt{SVMIX}, we further explain why \texttt{SGNN} works by clarifying the influence of \texttt{SGNN} on exploring the comprehensive optimal mapping from individual state values to the total state value through theoretical analysis. Moreover, through extensive simulations in two different scenarios, \texttt{SVMIX} manifests itself with a superior capability in terms of the performance like the convergence rate, the mean \& the variance of the average system utility and communication overheads. \par
However, the assumption that each agent can only make decisions based on the observation of itself might be over-emphasized as the communication between agents close to each other is allowed in practical scenarios in the decentralized execution. Moreover, a reward-decomposition method like \cite{rd} is able to sufficiently utilize the result of the centralized training during decentralized execution and even conduct decentralized training for fine tuning. In the future, it is meaningful to apply VD on the global reward in order to learn the individual reward functions and incorporate the inter-agent communication more comprehensively.

\bibliographystyle{IEEEtran}
\bibliography{reference}
\end{document}